\documentclass[aps,prb,twocolumn,superscriptaddress,showpacs]{revtex4-1}
\usepackage{graphicx,psfrag}
\usepackage{amsmath}

\begin{document}

\title{Anomalously slow spin dynamics and short-range correlations in the quantum spin ice systems Yb$_2$Ti$_2$O$_7$ and Yb$_2$Sn$_2$O$_7$} 

\author{A. Maisuradze}
\affiliation{Department of Physics, Tbilisi State University, Chavchavadze 3, GE-0128
Tbilisi, Georgia}
\affiliation{Laboratory for Muon-Spin Spectroscopy, Paul Scherrer Institute,
CH-5232 Villigen-PSI, Switzerland}
\author{P. Dalmas de R\'eotier}
\author{A. Yaouanc}
\affiliation{Universit\'e Grenoble Alpes, INAC-SPSMS, F-38000 Grenoble, France}
\affiliation{CEA, INAC-SPSMS, F-38000 Grenoble, France}
\author{A. Forget}
\affiliation{CEA, Centre de Saclay, DSM/IRAMIS/Service de Physique de l'Etat Condens\'e, F-91191 Gif-sur-Yvette, France}
\author{C. Baines}
\affiliation{Laboratory for Muon-Spin Spectroscopy, Paul Scherrer Institute,
CH-5232 Villigen-PSI, Switzerland}
\author{P.J.C. King}
\affiliation{ISIS Facility, Rutherford Appleton Laboratory, Chilton, Didcot, OX11 0QX, UK}

\date{\today}

\begin{abstract}
We report a positive muon spin relaxation and rotation ($\mu$SR) study of the quantum 
spin ice materials Yb$_2$Ti$_2$O$_7$ and Yb$_2$Sn$_2$O$_7$ focusing on the low field response. In agreement with earlier reports, data recorded in small longitudinal fields evidence anomalously slow spin dynamics in the microsecond range below the temperature $T_{\rm c}$ at which the specific heat displays an intense peak, namely $T_{\rm c}$ = 0.24~K and 0.15~K, respectively, for the two systems. We found that slow dynamics extends above $T_{\rm c}$ up to at least 0.7~K for both compounds. The conventional dynamical Gaussian Kubo-Toyabe model describes the $\mu$SR spectra recorded above $T_{\rm c}$. At lower temperatures a published analytical extension of 
the Gaussian Kubo-Toyabe model provides a good description, consistent with the existence of short-range
magnetic correlations. While the physical response of the two systems is qualitatively the same, Yb$_2$Ti$_2$O$_7$ exhibits a much larger local magnetic susceptibility than Yb$_2$Sn$_2$O$_7$ 
below $T_{\rm c}$. Considering previously reported ac susceptibility, neutron scattering and 
$\mu$SR results, we suggest the existence of anomalously slow spin dynamics to be a common physical property
of pyrochlore magnetic materials. The possibility of molecular spin substructures to be associated to the slow dynamics and therefore the short-range correlations is mentioned. The slow spin dynamics observed
under field does not exclude the presence of much faster dynamics detected in extremely low or zero field. 

\end{abstract}


\maketitle

\section{Introduction} 
\label{Introduction}

The pyrochlore insulator compounds $R_2M_2$O$_7$, where $R$ is a rare earth ion and $M$ a 
non magnetic element, have attracted considerable attention. The most remarkable discovery 
has been the spin-ice ground state of Ho$_2$Ti$_2$O$_7$ in 1997.\cite{Harris97} This ground 
state is now known for the emergence of a lattice-based Coulomb phase characterized 
by dipolar correlations and therefore pinch-point scattering patterns.\cite{Fennell09} 
Its excitations are emergent magnetic monopoles.\cite{Ryzhkin05,Castelnovo08,Morris09} 

This first success has led to systematic studies of other members of the $R_2M_2$O$_7$ 
family.\cite{Gardner10} However, apparently conflicting results have generated confusion.
Only recently clarifying experimental data and thorough analyses have appeared. For 
example, low temperature magnetization measurements for Tb$_2$Ti$_2$O$_7$ barely support
the existence of a magnetization plateau expected for a spin-ice like compound, at least
down to $0.02$~K.\cite{Yin13,Dalmas14a} A first physical explanation of the
persistent spin dynamics at low temperature as observed by the positive muon spectroscopy
($\mu$SR) spin-lattice relaxation rate has been proposed.\cite{Yaouanc15}
The dynamics would arise from low energy unidimensional excitations. They would be supported 
by spin loops originally discussed in Refs.~\onlinecite{Villain79,Lee02,Hermele04}.

Yb$_2$Ti$_2$O$_7$ stands alone. The originally proposed low temperature spin liquid state, 
derived from powder sample measurements,\cite{Hodges02,Yaouanc03} has been challenged ever since, with 
apparently no agreement reached on its basic physical properties, such as the absence or
existence of long-range magnetic correlations at low temperature as probed by neutron 
diffraction.\cite{Hodges02,Gardner04,Yasui03,Chang12} Remarkably, these correlations have 
recently been firmly established for a powder sample of the sibling compound 
Yb$_2$Sn$_2$O$_7$.\cite{Yaouanc13,Lago14}

One of the most mysterious physical property of magnetic pyrochlore compounds is the existence of anomalously slow spin dynamics. 
Here, a thorough study of the local magnetic fields and spin dynamics of Yb$_2$Ti$_2$O$_7$ 
and Yb$_2$Sn$_2$O$_7$ as probed by $\mu$SR spectroscopy under field is reported. Anomalously 
slow fluctuations of the Yb$^{3+}$ magnetic moments are detected below and above the 
temperature $T_{\rm c}$ at which specific heat displays a pronounced anomaly. Their effect on 
the measured spectra is fully described. Short-range correlations are detected below $T_{\rm c}$ for both compounds. Yb$_2$Ti$_2$O$_7$ is found to be characterized by a much 
larger local magnetic susceptibility below $T_{\rm c}$ than Yb$_2$Sn$_2$O$_7$. 

The organization of this paper is as follows. Section~\ref{Experimental_data_analysis} 
introduces the experimental method and explains the analysis of $\mu$SR spectra. 
In Secs.~\ref{Yb2Ti2O7_data} and \ref{Yb2Sn2O7_data} the Yb$_2$Ti$_2$O$_7$ and Yb$_2$Sn$_2$O$_7$ 
experimental spectra and their analysis are respectively presented. In the following section 
(Sec.~\ref{Discussion}) a discussion of the present data and previously reported extremely low-field results 
are provided. Conclusions are gathered in Sec.~\ref{Conclusions}. 

\section{Experimental and data analysis} 
\label{Experimental_data_analysis}

The measurements were performed on powder samples already used for different experimental investigations.\cite{Hodges02,Yaouanc03,Dalmas06a,Yaouanc11c,Yaouanc13,Yaouanc13a} Both of them 
are characterized by a well marked change in the longitudinal $\mu$SR signal at $T_{\rm c}$. 
For Yb$_2$Ti$_2$O$_7$ this is in agreement with the results of Ref.~\onlinecite{Chang14}, 
but in contrast to Ref.~\onlinecite{DOrtenzio13} for which a clear $\mu$SR signature of 
$T_{\rm c}$ only appears in the frequency shift. The magnetic transitions 
at $T_{\rm c}$ are first order with $T_{\rm c} \simeq 0.24$ and $0.15$~K for 
Yb$_2$Ti$_2$O$_7$ and Yb$_2$Sn$_2$O$_7$, respectively. This was first shown in Refs.~\onlinecite{Hodges02} 
and \onlinecite{Yaouanc13}, respectively. Later on, the first order nature
of the transition for Yb$_2$Ti$_2$O$_7$ was further characterized by detailed magnetization 
measurements.\cite{Lhotel14a} Hence, the magnetic history of a 
compound is expected to influence results of measurements in its ordered state. When 
presenting our spectra for $T < T_{\rm c}$ we shall explicitly describe the experimental 
conditions under which they were recorded. Surprisingly enough, for both compounds no spontaneous
$\mu$SR magnetic field has been observed below $T_{\rm c}$, even for a powder sample of 
Yb$_2$Sn$_2$O$_7$ for which magnetic Bragg reflections have been detected as mentioned in the 
introduction.

The $\mu$SR measurements were performed at the MuSR spectrometer of the ISIS facility 
(Rutherford Appleton Laboratory) and the LTF spectrometer of the Swiss Muon Source (S$\mu$S, Paul Scherrer Institute).

The measurements were mostly done with the longitudinal geometry for which the external field
${\bf B}_{\rm ext}$ is set along the initial muon beam polarization ${\bf S}_\mu$.\cite{Yaouanc11} 
In fact, it occurs that ${\bf B}_{\rm ext}$ and ${\bf S}_\mu$ are not strictly parallel due to the presence of a separator in the beamlines. This is particularly true at LTF.\footnote{In the longitudinal geometry, the beamline spin rotator is supplied with low voltage and current to act as a separator, namely a device used to remove unwanted particles, e.g.\ electrons, from the beam. See http://www.psi.ch/smus/pim3.} 
Hence, while the longitudinal polarization function derives from the counts detected in the detectors placed parallel and antiparallel to ${\bf B}_{\rm ext}$, an oscillating signal is observed in the detectors set perpendicular to ${\bf B}_{\rm ext}$. Although its amplitude is small it contains valuable information which we have used in this study. The results obtained in this way were confirmed by recording a few spectra in the usual transverse geometry, i.e.\ with ${\bf S}_\mu$ oriented perpendicular to ${\bf B}_{\rm ext}$.\footnote{In fact, at LTF the angle between ${\bf S}_\mu$ and ${\bf B}_{\rm ext}$ is about 57$^\circ$.}

We now present the framework for the data analysis.
The polarization function probing a static magnetic field distribution with field component distributions $D_{\rm c}(B^\alpha)$ identical for the three Cartesian directions $\alpha=X$, $Y$, or $Z$, is derived from the Larmor equation. In the longitudinal geometry, it is expressed as
\begin{widetext}
\begin{equation}
P_{Z}^{\rm stat}(t) = {1 \over {\mathcal Z}} \iiint_{-\infty}^{\infty}   
\left\{ \left({B^{Z}\over B}\right)^2 + \left[1-\left({B^{Z}\over B}\right)^2 \right]
\cos(\omega_\mu t )  \right\} D_{\rm c}(B^X)  D_{\rm c}(B^Y) 
D_{\rm c}\left({ B^Z - B_{\rm long} \over {\mathcal Z}}\right) {\rm d} B^X {\rm d} B^Y  {\rm d} B^Z,
 \label{eq:asyStatDef}
\end{equation}
\end{widetext}
with $B^2 = (B^X)^2 + (B^Y)^2 + (B^Z)^2$, $\omega_\mu = \gamma_\mu B$ where $\gamma_\mu$ is the muon
gyromagnetic ratio ($\gamma_\mu=2\pi\times 135.53 \times 10^6 \, {\rm rad} \, {\rm s}^{-1} {\rm T}^{-1}$), and
${B}_{\rm long}$ is the longitudinal field along the $Z$-axis which may not be necessarily 
equal to $B_{\rm ext}$. The parameter ${\mathcal Z}$ describes a possible uniaxial anisotropy in the $Z$ direction.  
Unless ${\mathcal Z}$ is specified, we shall assume ${\mathcal Z} = 1$, i.e.\ no anisotropy.

In fact the field at the muon 
sites has often a dynamical character. The muon polarization 
function $P_Z(t)$ for a magnetic field fluctuating at rate $\nu_{\rm c}$ is computed using the strong collision model.\cite{Kehr78,Yaouanc11} It satisfies the following integral equation:
\begin{align}
P_Z(t) =& P_Z^{\rm stat}(t)\exp(-\nu_{\rm c}t) + \nonumber \\
& \nu_{\rm c} \int^t_0 P_Z (t - t^\prime) P_Z^{\rm stat} (t^\prime) \exp(-\nu_{\rm c} t^\prime)
{\rm d} t^\prime.
\label{eq:dynamics_1}
\end{align}

To proceed further an expression for  $D_{\rm c}(B^\alpha)$ is needed. A conventional choice is a Gaussian function. 
Understanding $P_Z(t)$ measured for Yb$_2$Ti$_2$O$_7$ below $T_{\rm c}$ required the following simple extension of the Gaussian shape:\cite{Yaouanc13a}
\begin{align}
 D_{\rm c}(x \delta) = {1\over N \delta }
 \exp\left[-{1\over 2}x^2 - {1\over 3} (\eta_3 x)^3 - {1\over 4}(\eta_4 x)^4 \right].
\label{eq:FldDistrModelPRB}
\end{align}
Here, $\delta$ sets the magnetic field scale and $N$ is the 
normalization constant. In zero field and when $\eta_3=\eta_4=0$, i.e.\ for a Gaussian field
component distribution, $P_Z^{\rm stat}(t)$ is given by the famous Kubo-Toyabe formula.\cite{Hayano79} 
Under field and still in the Gaussian limit and with ${\mathcal Z} = 1$, the three-dimensional integral in 
Eq.~\ref{eq:asyStatDef} reduces to a one-dimensional integral.\cite{Yaouanc11} In zero 
field only symmetrized distributions are probed.\cite{Dalmas14} 
In principle, in longitudinal field, $\mu$SR can resolve asymmetric distributions. However, concerning the data presented in this report, no clear deviation from symmetric distributions was found in a first analysis.\footnote{Note that, in addition, we expect the field component distribution measured on a powder to be symmetric, provided there is no field-induced anisotropy.} Therefore, the distributions entering Eq.~\ref{eq:asyStatDef} have always been symmetrized distributions $D_{\rm c}^{\rm sym}(B^\alpha)$ defined as
\begin{align}
D_{\rm c}^{\rm sym}(B^\alpha) = {1\over 2}\left[D_c(B^\alpha) +  D_c(-B^\alpha)\right]. 
\label{eq:FldDistrModelPRB_sym}
\end{align}
Once $D_{\rm c}^{\rm sym}(B^\alpha)$ is determined, the standard deviation $\Delta_{\rm LF}$ of the distribution can be computed numerically according to
\begin{align}
\Delta_{\rm LF}^2 = \int_{-\infty}^{\infty} b^2 D_{\rm c}^{\rm sym} (b) {\rm d}b,
\end{align}
since the average of $D_{\rm c}^{\rm sym} (b)$ is 0. If $\eta_3 = \eta_4 = 0$, $\Delta_{\rm LF} = \delta$.

Equation~\ref{eq:FldDistrModelPRB} is an ansatz which assumes the field component distribution 
to deviate only slightly from the Gaussian shape. It can account for short-range spin-correlation 
effects.\cite{Yaouanc13a} The presence of short-range correlations is already known to transform the Kubo-Toyabe function into a polarization function with a shallower minimum.\cite{Noakes99} It is natural to apply Eq.~\ref{eq:FldDistrModelPRB} for the description of 
longitudinal field (LF) $\mu$SR asymmetry spectra recorded in a paramagnetic state. It can 
even be used below $T_{\rm c}$ when a spontaneous field is not detected, e.g.\ because of its 
dynamical nature.\cite{Dalmas06} 
 
LF spectra will be analyzed with the two-component model
\begin{equation}
 A^{\rm LF}(t) = A^{\rm LF}_0 \left[ (1-F_{\rm bg})P_Z(t) + F_{\rm bg}\right],
 \label{eq:AsyModel:LF}
\end{equation}
where $A^{\rm LF}_0$ is the initial $\mu$SR asymmetry, $F_{\rm bg}$ the fraction of muons 
stopped outside of the sample, for which the relaxation is negligible.

Except for some spectra recorded in the ordered state of Yb$_2$Ti$_2$O$_7$, asymmetry time spectra corresponding to 
the transverse field (TF) geometry will be described with a weighted sum of two Gaussian damped oscillations:
\begin{align}
& A^{\rm TF}(t) = \nonumber \\ 
& A^{\rm TF}_0  \left[ (1-F_2) \exp\left(-{\gamma_\mu^2\Delta_{{\rm TF},1}^2t^2 \over 2}\right)
\cos(\gamma_\mu B_1t+\phi) \right.\nonumber \\
& + \left. F_2\exp\left(-{\gamma^2_\mu \Delta_{{\rm TF},2}^2t^2 \over 2}\right)
\cos(\gamma_\mu B_2t+\phi)\right]. 
\label{eq:TFmodel}
\end{align}
Here, $\phi$ is an experimental phase parameter and $A_0^{\rm TF}$ the initial asymmetry 
in the transverse geometry. For the vast majority of measurements for which LF and TF spectra
were simultaneously recorded at LTF, $A_0^{\rm TF} \simeq 0.06$. The relation 
$(A_0^{\rm LF})^2 + (A_0^{\rm TF})^2 = 0.27^2$ holds since the total asymmetry is a 
characteristics of the spectrometer. 
The mean fields $B_1$, $B_2$, and the standard deviations $\Delta_{{\rm TF},1}$ and 
$\Delta_{{\rm TF},2}$ refer to the first and second components, respectively. 
We find $\Delta_{{\rm TF},2}\ll \Delta_{{\rm TF},1}$,  $F_2 = F_{\rm bg}$ within 
experimental uncertainties, and $B_2 \simeq B_{\rm ext}$. These results are consistent with
the attribution of the second component to the background, i.e.\ to muons missing the sample. 
Its narrow signal can serve as a precise measurement of $B_{\rm ext}$.
For the first component we resorted to a simple function rather than trying to take into account 
a possible deviation from the Gaussian field component distribution.\cite{Yaouanc13a} This is 
justified since, while we shall find deviations from the conventional description for
the longitudinal asymmetries at low temperature, the simple expression in Eq.~\ref{eq:TFmodel} 
provides a proper description of the measured transverse-field asymmetries.

\section{Experimental results: Y\lowercase{b}$_2$T\lowercase{i}$_2$O$_7$ } 
\label{Yb2Ti2O7_data}

We shall first consider spectra recorded in the paramagnetic phase, i.e.\ above $T_{\rm c} \simeq 0.24$~K. In a second step data taken below $T_{\rm c}$ will be presented.

\subsection{Paramagnetic state} 
\label{Yb2Ti2O7_data_para}

\begin{figure}[t]
\hspace{-0.8cm}
\includegraphics[width=0.50\linewidth ]{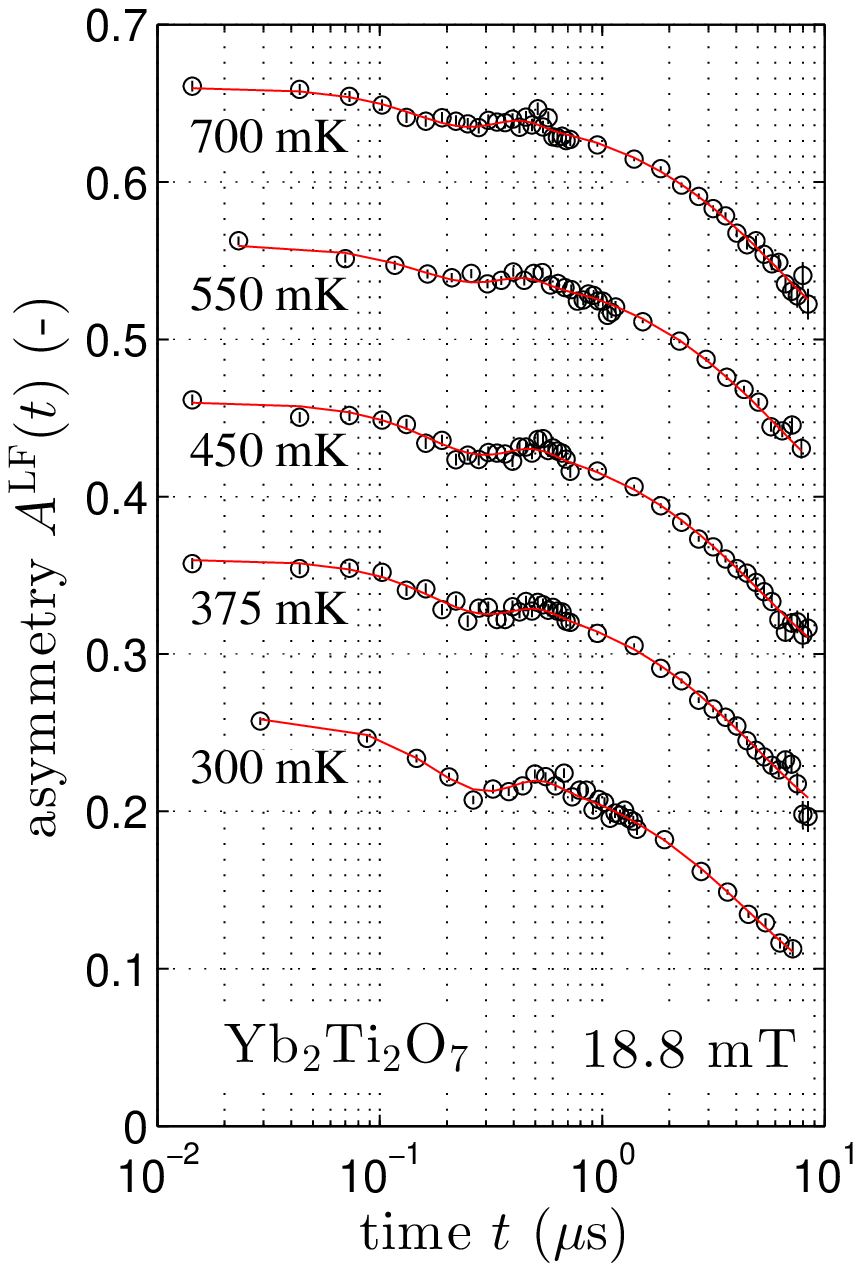}
\includegraphics[ width=0.50\linewidth]{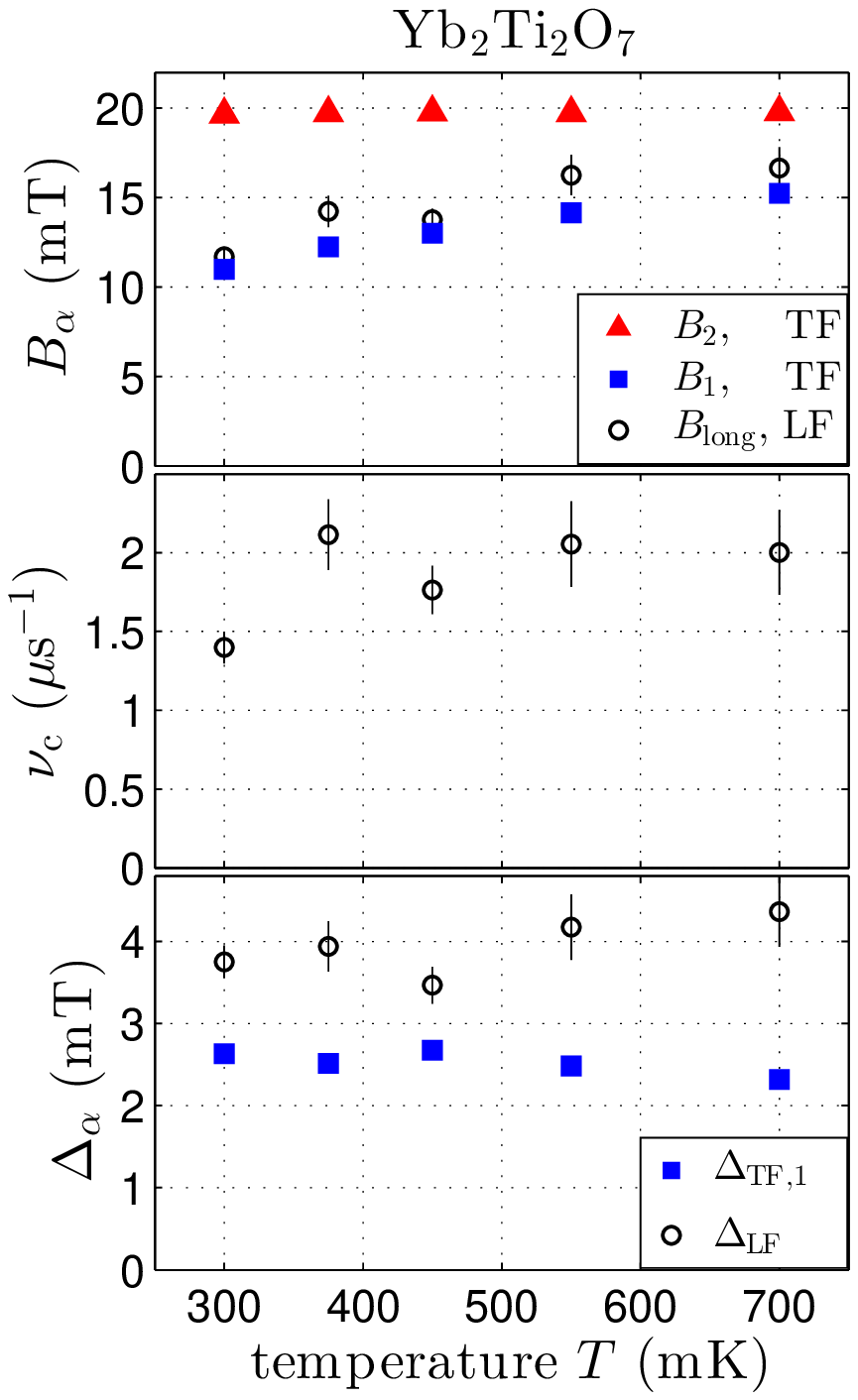}
\hspace{-0.8cm}
\caption{(color online). $\mu$SR asymmetry spectra recorded at LTF in the paramagnetic phase of a Yb$_2$Ti$_2$O$_7$ powder sample under $B_{\rm ext} = 18.8$~mT. 
(left) Temperature scan of the longitudinal asymmetry. Circles are experimental data while 
solid lines represent fits as explained in the main text. The spectra for consecutive temperatures are vertically shifted by 0.1 for better visualization. A logarithmic time scale 
is used. (right) Thermal dependence of fitting parameters extracted for spectra recorded in the longitudinal and transverse geometries.}
\label{Yb2Ti2O7_para_Tscan_spectra}
\end{figure}

We first show in Fig.~\ref{Yb2Ti2O7_para_Tscan_spectra}
asymmetry spectra recorded with a longitudinal field $B_{\rm ext} = 18.8$~mT at different temperatures $T > T_{\rm c}$. 
Remarkably, the asymmetry does not decay monotonically, but displays a well defined local minimum, at about 0.25~$\mu$s at low temperature. The structure is less pronounced as the 
sample is warmed, but is still discernible at $700$~mK, pointing to a relatively slow
spin dynamics in the paramagnetic state.\cite{Yaouanc11} The spectra were analyzed with 
Eqs.~\ref{eq:asyStatDef}--\ref{eq:AsyModel:LF}. The conventional dynamical Gaussian Kubo-Toyabe model was 
found to provide a good description, i.e.\ $\eta_3= \eta_4 = 0$. 

\begin{figure}[ht]
\includegraphics[width=0.49\linewidth]{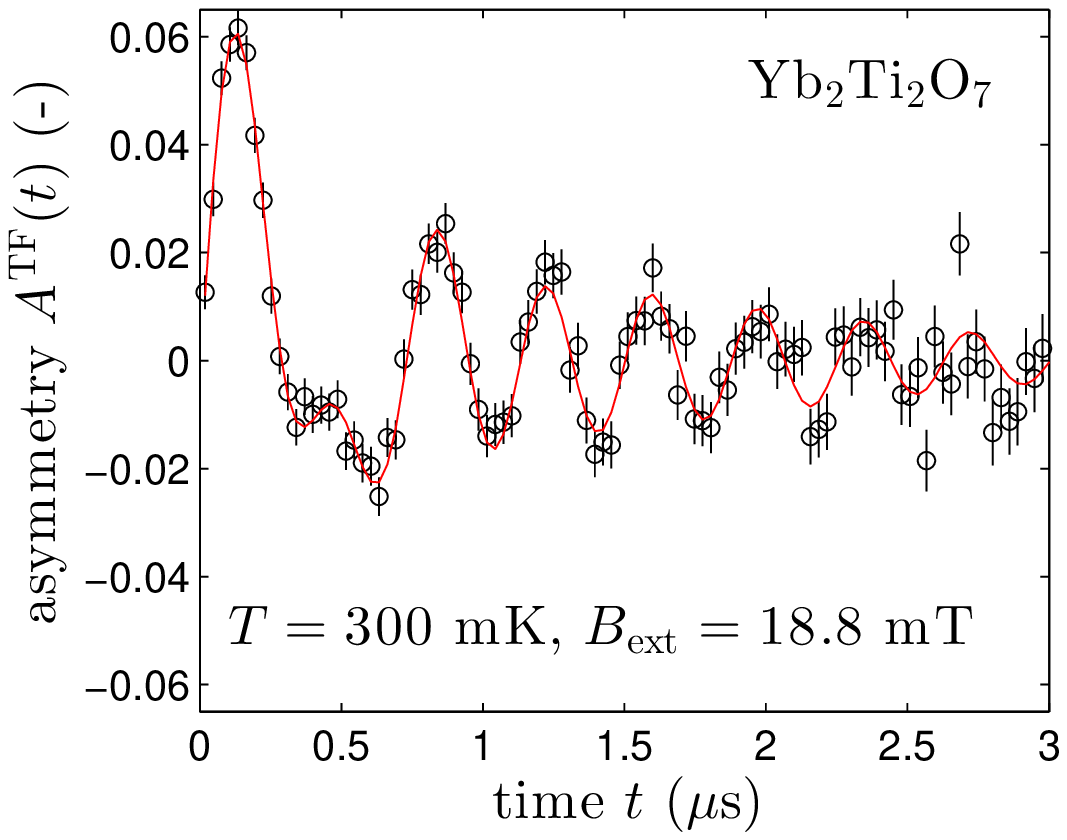}    
\includegraphics[width=0.49\linewidth]{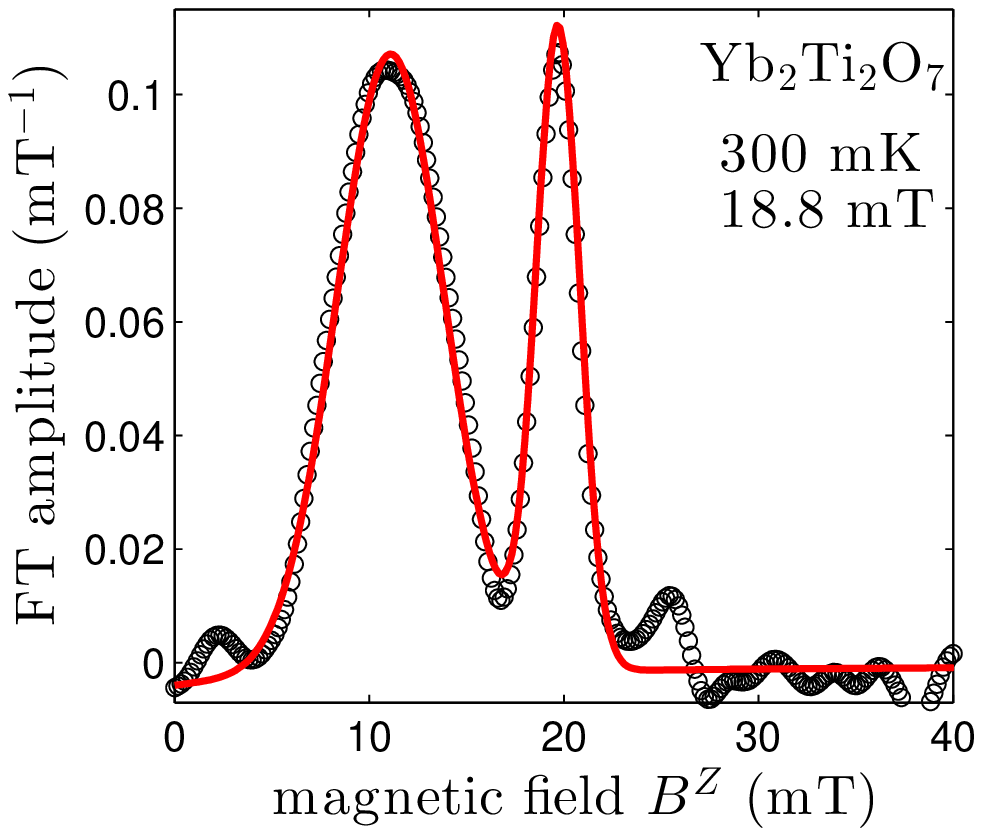}
\caption{(color online). (left) Example of a transverse-field asymmetry spectrum measured at 300~mK 
under $B_{\rm ext} = 18.8$~mT for a powder sample of Yb$_2$Ti$_2$O$_7$. The spectrum was recorded at 
LTF with the spectrometer in the longitudinal geometry. This explains the small value for the asymmetry.
Black circles display the spectrum while the 
red curve results from a fit to the data with Eq.~\ref{eq:TFmodel} as  explained in the main text. 
(right) The corresponding Fourier transform amplitude of the asymmetry data and fitting curve. 
}
\label{Yb2Ti2O7_FTexample}
\end{figure}

Transverse geometry spectra were analyzed with Eq.~\ref{eq:TFmodel} and an example 
of such analysis is shown in the left panel of Fig.~\ref{Yb2Ti2O7_FTexample}. 
In the right panel the Fourier transform (FT) amplitude of the spectrum is displayed. Only two peaks are resolved, in contrast to 
Ref.~\onlinecite{DOrtenzio13} where three peaks --- one broad and two narrow --- are reported for a powder sample, both above and below $T_{\rm c}$. This will be further discussed at 
the end of Sec.~\ref{Yb2Ti2O7_data_order}.

The thermal dependences of the parameters extracted from the LF and TF analyses are presented in 
the three panels on the right of Fig.~\ref{Yb2Ti2O7_para_Tscan_spectra}. As already noticed, 
$B_2$ is a good measure of $B_{\rm ext}$. One can see that both $B_1$ and $B_{\rm long}$ are 
closely related and are substantially smaller than $B_2 \simeq B_{\rm ext}$. These important results will be discussed in Sec.~\ref{Discussion_LF}.
On increasing temperature this reduction is attenuated. The fluctuation rate 
of the spin dynamics is found to be $\nu_{\rm c}\simeq 2$~$\mu$s$^{-1}$ confirming the  
anomalously slow spin dynamics already inferred from visual examination of the spectra. 
We find $\Delta_{\rm LF}$ and $\Delta_{{\rm TF},1}$ to be roughly temperature independent.
In contrast to expectation, these parameters are quite different. However, we note that taking ${\mathcal Z=0.8}$, the fit of the spectra are of similar quality and within uncertainties we still get the same results for $B_\alpha$ and $\nu_{\rm c}$, but now $\Delta_{\rm LF}$ and $\Delta_{{\rm TF},1}$ are approximately equal. The anisotropy of the field component distribution that this result suggests, could be a field-induced effect. The important point is that $B_1$ and $B_{\rm long}$ remains nearly equal. 

\begin{figure}
\hspace{-0.8cm}
\includegraphics[width=0.49\linewidth]{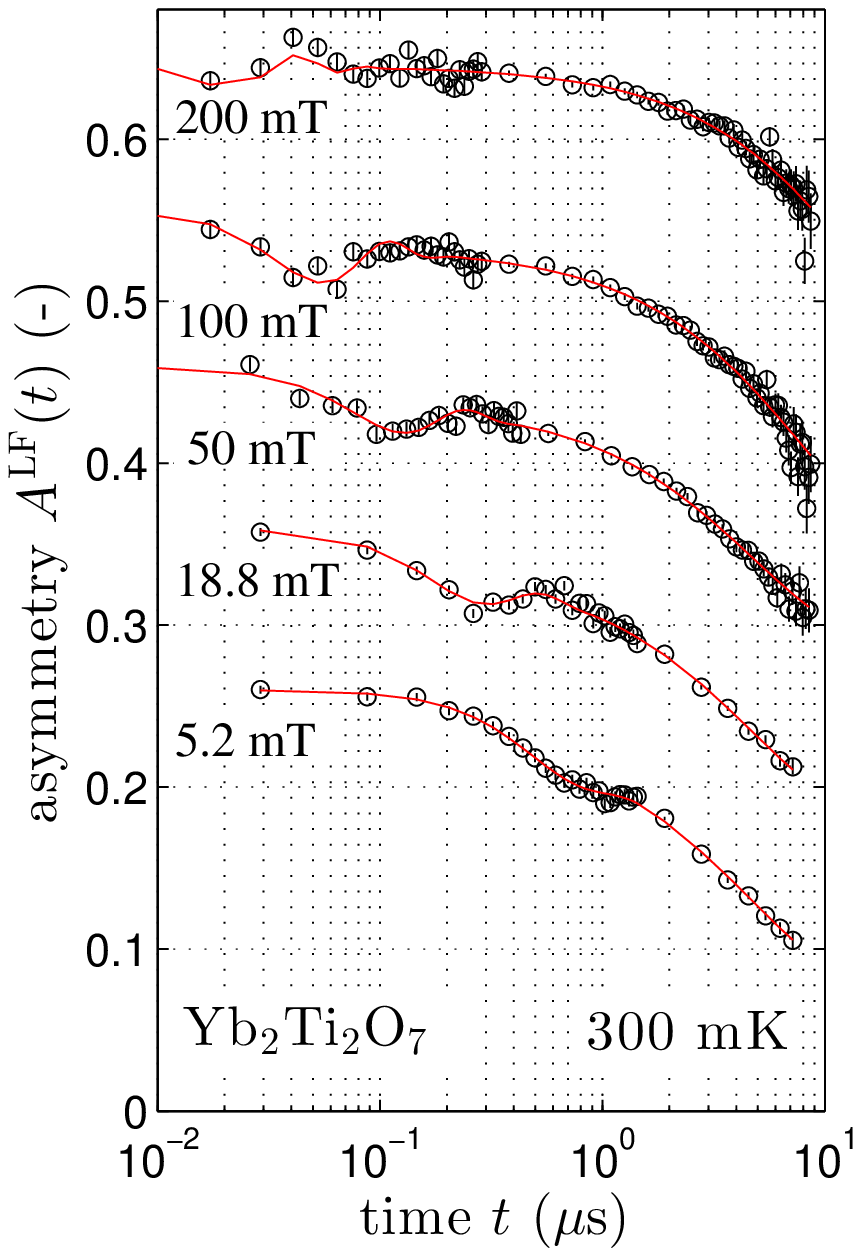}
\includegraphics[width=0.49\linewidth]{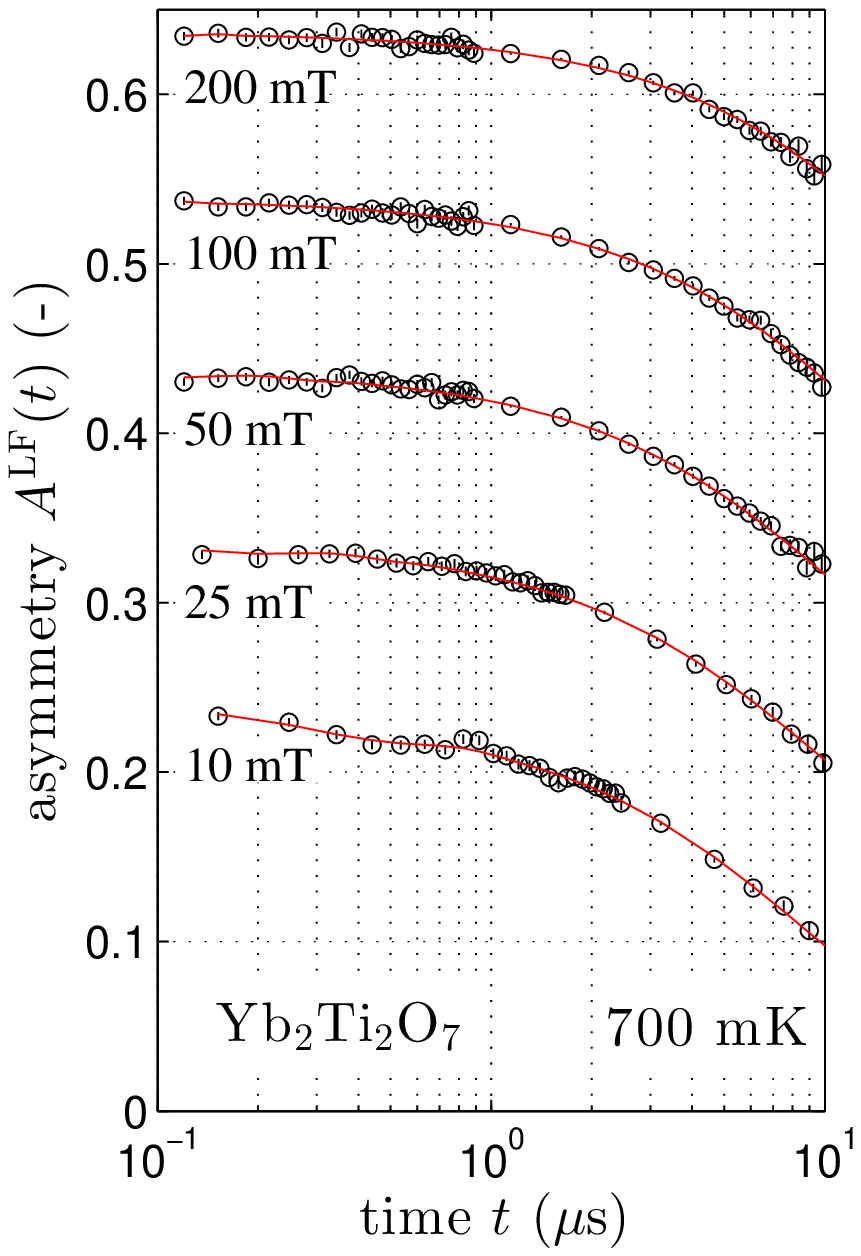}
\hspace{-0.8cm}
\caption{(color online). Field dependence of asymmetry spectra recorded 
in the paramagnetic phase of a Yb$_2$Ti$_2$O$_7$ powder sample in the longitudinal geometry. 
The data are represented by black circles and the corresponding fitting curves as solid lines.
The fits are explained in the main text. The spectra were recorded at LTF for $T = 300$~mK (left) 
and at MuSR for $T= 700$~mK (right). The data for consecutive $B_{\rm ext}$ values (indicated nearby each spectrum) are shifted by 0.1 unit for better visualization. 
}
\label{Yb2Ti2O7_para_Fscan_spectra}
\end{figure}

\begin{figure}
\centering
\includegraphics[width=0.50\linewidth]{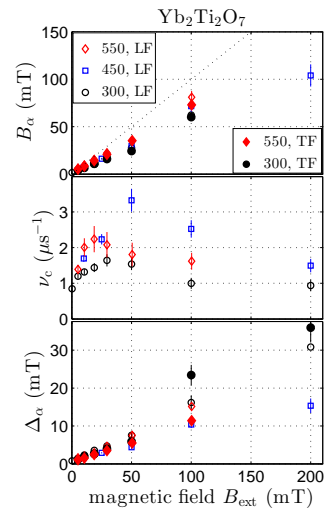}
\caption{(color online). External magnetic field dependence of physical parameters extracted from 
the analysis of field scans performed in the paramagnetic phase of a Yb$_2$Ti$_2$O$_7$ powder
sample. We first plot $B_{\rm long}$ and $B_1$ (collectively denoted as $B_\alpha$), then 
$\nu_{\rm c}$, and finally $\Delta_{\rm LF}$ and $\Delta_{{\rm TF},1}$ (collectively denoted as 
$\Delta_\alpha$).
Most of the data have been derived from LF spectra (empty symbols), with few results obtained with the TF 
(full symbols) geometry at 300 and 550~mK. The plotting symbols correspond to different temperatures (given in units of millikelvin) and experimental geometries as indicated in the insets. 
}
\label{Yb2Ti2O7_para_parameters}
\end{figure}

In order to confirm the unusual slow spin dynamics we recorded data for a set of longitudinal fields at 300, 450, 550, 
and 700~mK, i.e.\ above $T_{\rm c}$. Examples of spectra and fitting curves are presented in 
Fig.~\ref{Yb2Ti2O7_para_Fscan_spectra}. 
Again the dynamical Gaussian Kubo-Toyabe model gives a nice description of the spectra. Field-induced 
oscillations are clearly observed in a broad field range, 
supporting the existence of slow fluctuations. The fit parameters are plotted in Fig.~\ref{Yb2Ti2O7_para_parameters} as a function of $B_{\rm ext}$ for temperatures up to 550~mK. 
The oscillations amplitude being somewhat reduced when the temperature is increased, some of the fit parameters are indeed poorly defined at higher temperatures. The reduction of the field probed by the muons is confirmed;
see the plot of $B_\alpha$ 
versus $B_{\rm ext}$. 
The parameters $\Delta_{\rm LF}$ and $\Delta_{{\rm TF},1}$ characterizing the standard deviation of the Gaussian 
field distribution in LF and TF geometries, respectively, increase roughly linearly with $B_{\rm ext}$. 

\subsection{Ordered magnetic state} 
\label{Yb2Ti2O7_data_order}

In a previous work it was shown that understanding a spectrum recorded at $200$~mK, i.e.\ below $T_{\rm c}$, and at low field for Yb$_2$Ti$_2$O$_7$ required an extension of the conventional Gaussian distribution model.\cite{Yaouanc13a} This was confirmed with a numerical method based on the maximum
entropy principle and the reverse Monte Carlo algorithm.\cite{Dalmas14} 
\begin{figure}[!t] 
\begin{picture}(245,202)
\put(-8,90){\includegraphics[width=0.52\linewidth]{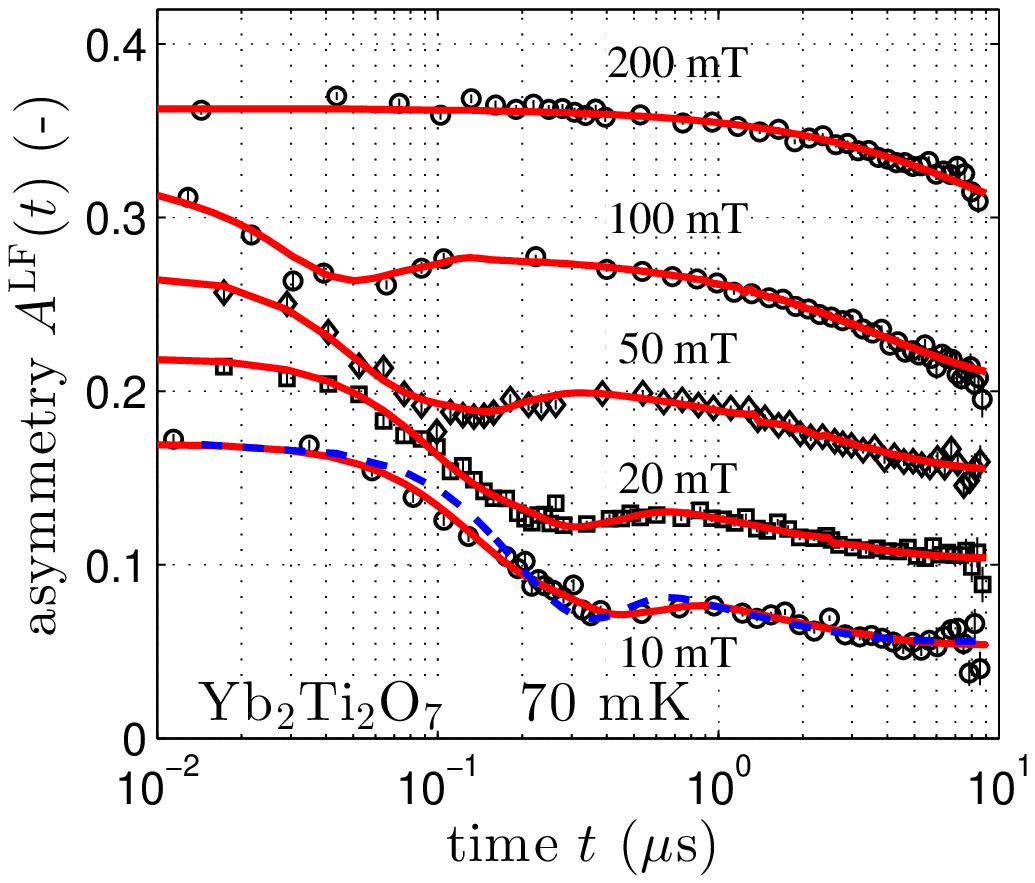}}
\put(97,113){\small(a)}
\put(5,0){\includegraphics[width=0.45\linewidth]{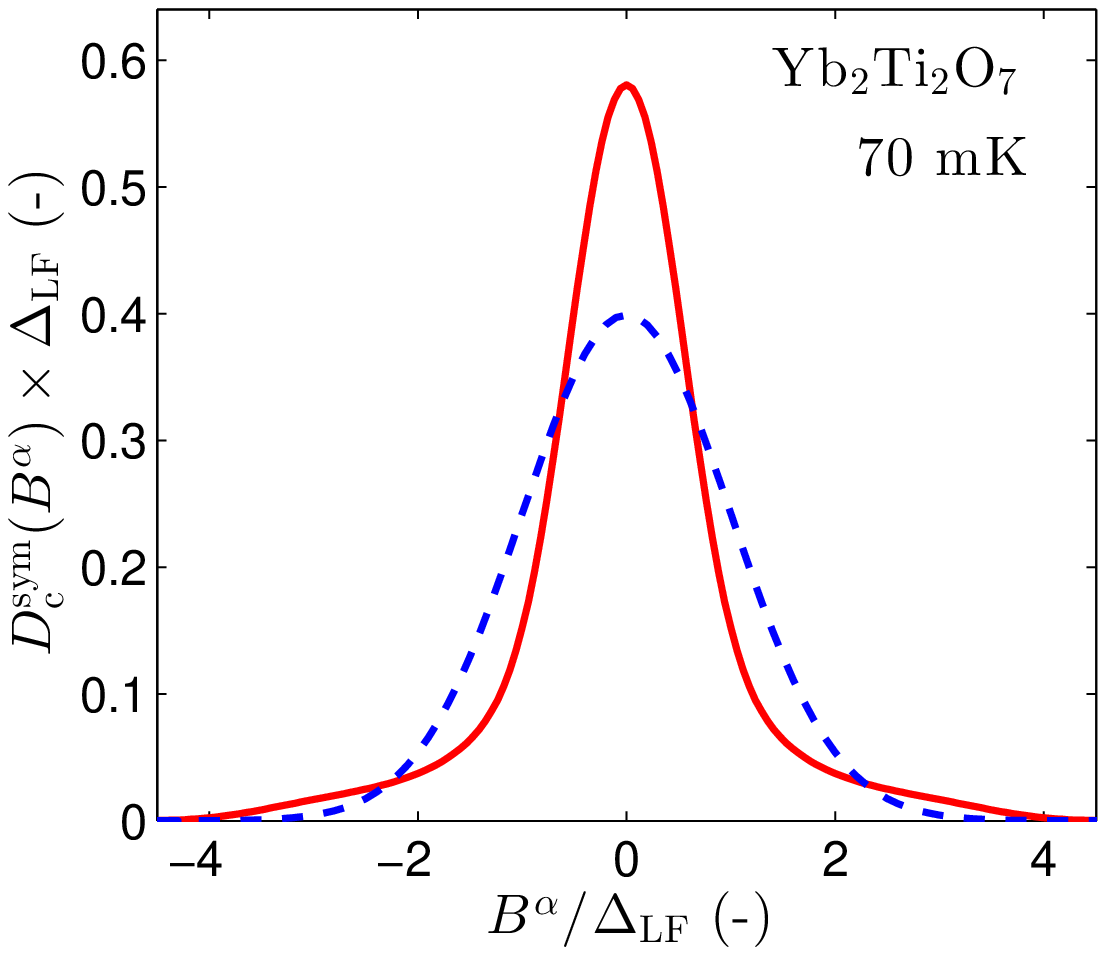}}
\put(97,30){\small(e)}
\put(122,10){\includegraphics[width=0.45\linewidth]{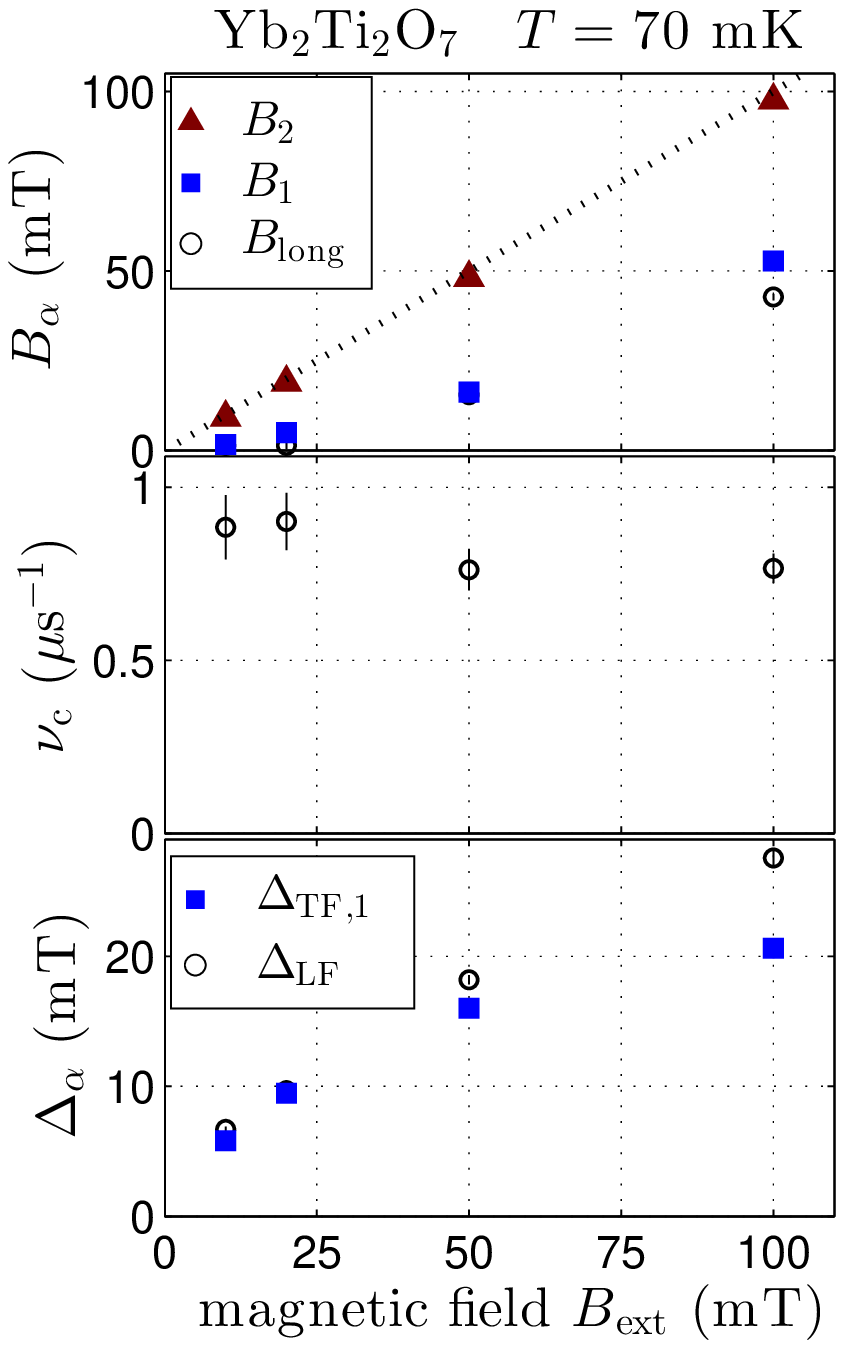}}
\put(215,129){\small(b)}
\put(215,83){\small(c)}
\put(215,33){\small(d)}
\end{picture}
\caption{(color online). Data recorded for a powder sample of Yb$_2$Ti$_2$O$_7$ at $T =70$~mK with the longitudinal geometry. (a) Measured asymmetries with increasing $B_{\rm ext}$ indicated nearby each spectrum. 
The spectra are vertically shifted by 0.05 for better visualization. The solid lines result from analytical fits as 
explained in the main text. For comparison, the blue dashed line represents the dynamical Gaussian Kubo-Toyabe fit of the asymmetry measured in a field of 10~mT. (b, c, d) Empty circles show the field dependence of three parameters describing the asymmetry spectra: the field 
$B_{\rm long}$, the fluctuation rate $\nu_{\rm c}$ and the standard deviation $\Delta_{\rm LF}=1.70 \, \delta$ for $\eta_3=0.7335$ and $\eta_4=0.4558$. Full symbols represent the field dependence of $B_1$, $B_2$, and $\Delta_{\rm TF,1}$ obtained from the TF spectra (not shown). 
The dotted line in panel (b) corresponds to the curve $B_{\rm long}=B_{\rm ext}$. (e) Field component distribution in reduced field scale. The solid line results from the analysis with the extension of the Gaussian model. The dashed line is for the Gaussian field distribution corresponding to the dynamical Gaussian Kubo-Toyabe fit in panel (a). 
}
\label{Yb2Ti2O7_ordered_Fscan_spectra}
\end{figure}
In  Fig.~\ref{Yb2Ti2O7_ordered_Fscan_spectra}(a) 
we display spectra recorded for different longitudinal fields at $70$~mK, i.e.\ below $T_{\rm c}$, at LTF.
Spectra taken with the
same experimental conditions at MuSR were published elsewhere.\cite{Dalmas06a} Here we profit from the better time resolution available 
at LTF. The sample was zero-field cooled. The data were subsequently recorded for increasing values of the field. 
Weak oscillations are 
observed, with an amplitude which is again reduced at high field. The asymmetry spectra were analyzed with Eqs.~\ref{eq:asyStatDef}--\ref{eq:AsyModel:LF}. The separate analysis of the spectra led to parameters $\eta_3$ and $\eta_4$ independent of the field, within uncertainty. Therefore in a second step, the spectra at different fields were simultaneously fit with the same set of parameters $\eta_3$ and $\eta_4$. The fits shown in Fig.~\ref{Yb2Ti2O7_ordered_Fscan_spectra}(a) are excellent.
We found $\eta_3 = 0.73 \, (2)$ and 
$\eta_4 = 0.46 \, (2)$. This is consistent with the previous analysis of a spectrum recorded
at $200$~mK under $2$~mT.\cite{Yaouanc13a} The two parameters $\eta_3$ and $\eta_4$ are not negligible. 
This means that $D_{\rm c}^{\rm sym}(B^\alpha)$ displayed in Fig.~\ref{Yb2Ti2O7_ordered_Fscan_spectra}(e) deviates 
substantially from the conventional Gaussian shape as the comparison shows. From the values of $\eta_3$ and $\eta_4$ we compute the field distribution standard deviation $\Delta_{\rm LF}$ = 1.70\,(9)\,$\delta$. Note that the parameters $\eta_3$ and $\eta_4$ are strongly correlated.
The inferred $D_{\rm c}^{\rm sym}(B^\alpha)$ is characterized by relatively pronounced tails. The fitting parameters $B_{\rm long}$, 
$\nu_{\rm c}$, and $\Delta_{\rm LF}$ are presented in 
Figs.~\ref{Yb2Ti2O7_ordered_Fscan_spectra}(b)--\ref{Yb2Ti2O7_ordered_Fscan_spectra}(d).  We find 
$B_{\rm long} \ll B_{\rm ext}$ with $B_{\rm long}$ roughly proportional to $B_{\rm ext}$, as in the 
paramagnetic state but with a much smaller coefficient of proportionality. The fluctuation 
rate $\nu_{\rm c}\lesssim 1 \, \mu {\rm s}^{-1}$ is smaller than in the paramagnetic state but still 
appreciable. The standard deviation $\Delta_{\rm LF}$ increases quite linearly with field, in agreement with the behavior in the 
paramagnetic state. $\Delta_{{\rm TF},1}$ is slightly  different from $\Delta_{\rm LF}$ at high field. As for the paramagnetic state,
this difference becomes negligible if a small field-induced anisotropy is introduced. This is a side effect which
can be neglected, as already mentioned when the paramagnetic data were discussed.

The strong reduction of $B_{\rm long}$ relative to $B_{\rm ext}$ is remarkable. As expected, and supporting our data analysis,
$B_1 \simeq B_{\rm long}$. We expand on the meaning of this reduction in Sec.~\ref{Discussion_LF}.

\begin{figure}[t]
\includegraphics[width=0.49\linewidth]{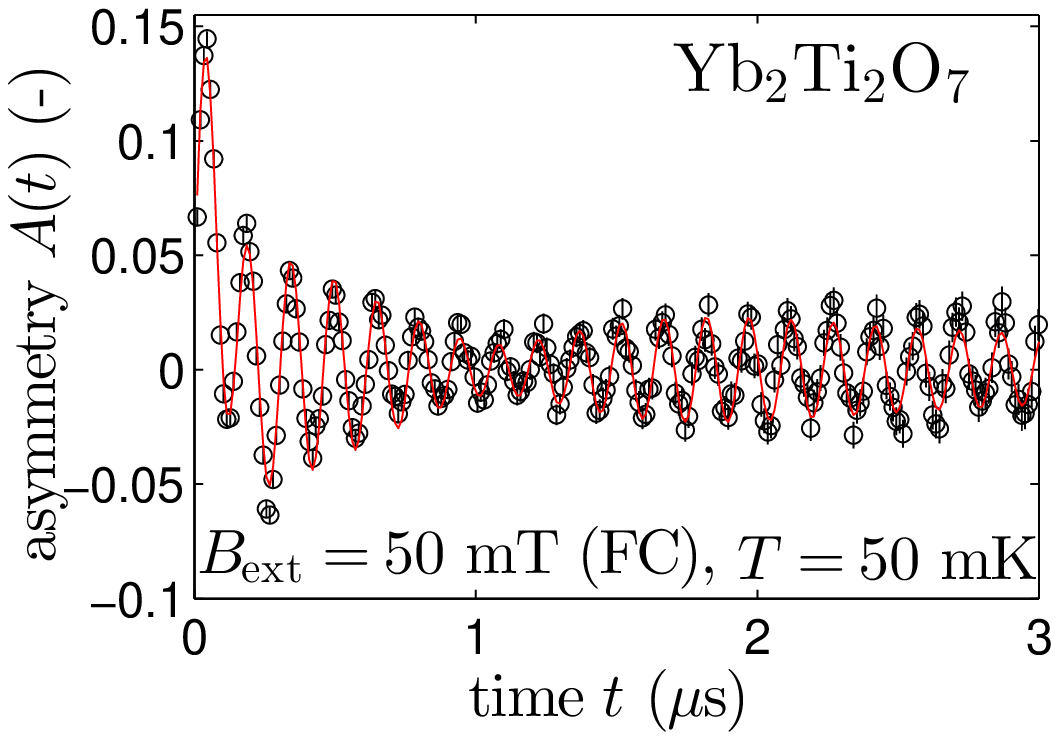}
\includegraphics[width=0.49\linewidth]{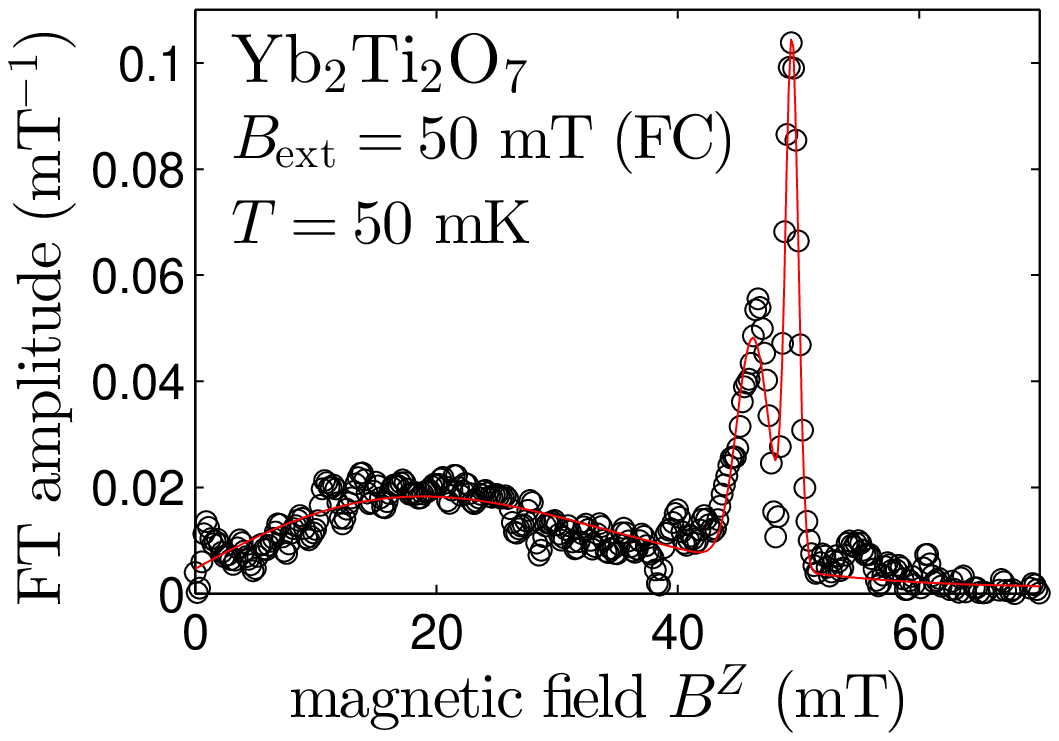}  
\caption{(color online). (left) Example of a transverse-field asymmetry spectrum measured at 50~mK under $B_{\rm ext} = 50$~mT in 
field cooling condition for a powder sample of Yb$_2$Ti$_2$O$_7$. The spectrum was recorded at LTF with the spectrometer in the 
transverse geometry. Black circles display the spectrum while the red curve results from a fit with a model similar to 
Eq.~\ref{eq:TFmodel} but with three Gaussian oscillating components. (right) The corresponding Fourier transform amplitude of the asymmetry data 
and fitting curve. The broad signal stems from the sample while two narrow peaks at $B^Z\simeq 46$~mT and 
$B^Z\simeq 49$~mT originate from muons stopped outside the sample. 
}
\label{Yb2Ti2O7_FTexample:ordered}
\end{figure}

In order to resolve the discrepancy between our results and those reported in 
Ref.~\onlinecite{DOrtenzio13}, we performed additional measurements at the LTF spectrometer 
in TF configuration. 
This allows us to simultaneously record LF and TF spectra with a large $A_0^{\rm TF}=0.21$, and a smaller $A_0^{\rm LF}=0.17$.
The experimental conditions, namely measurements in field cooling protocol  with $B_{\rm ext}=50$~mT and at $T=50$~mK, are as in 
Ref.~\onlinecite{DOrtenzio13}. The result is shown in Fig. \ref{Yb2Ti2O7_FTexample:ordered}. 
Qualitatively our Fourier spectrum is consistent with that reported in Ref.~\onlinecite{DOrtenzio13}. Three peaks are visible. 
This is in contrast to the spectrum displayed in Fig. \ref{Yb2Ti2O7_FTexample} for which only two peaks are present. Regarding this latter
spectrum, $B_{\rm ext}$ is smaller and the data concern the paramagnetic instead of the ordered state. While it is natural to assign the broad peak around 20~mT in Fig. \ref{Yb2Ti2O7_FTexample:ordered} to the sample, we believe that the other two arise from muons stopped outside the sample. Indeed, the relative contributions of the peaks at 46 and 49~mT to the Fourier spectrum are 14\,(1)\% and 12\,(1)\%, respectively. 
Assigning only the peak at 49~mT to the background leads to an unreasonably small background value for the LTF spectrometer. Most probably, the stray field and the geometry of the sample and sample-holder are at the origin
of the two peaks on the right of the field scale. The narrow peak should correspond to the part of the sample-holder located far 
from the sample while a slightly broader peak originates from the sample-holder part in the immediate neighborhood to the sample.  

In any case, the agreement demonstrated in Fig.~\ref{Yb2Ti2O7_ordered_Fscan_spectra}(b) between the mean fields $B_1$, measured
directly from TF spectra, and $B_{\rm long}$, extracted from the analysis of LF spectra, is a strong support for the methodology
of our TF and LF data analysis.

\section{Experimental results: Y\lowercase{b}$_2$S\lowercase{n}$_2$O$_7$} 
\label{Yb2Sn2O7_data}

Here we present results of the analysis for Yb$_2$Sn$_2$O$_7$ in the same way as was done for 
Yb$_2$Ti$_2$O$_7$. We first focus on the paramagnetic state.
 
\subsection{Paramagnetic state} 
\label{Yb2Sn2O7_data_para}

\begin{figure}[!t]
\hspace{-0.8cm}
\includegraphics[width=0.5\linewidth ]{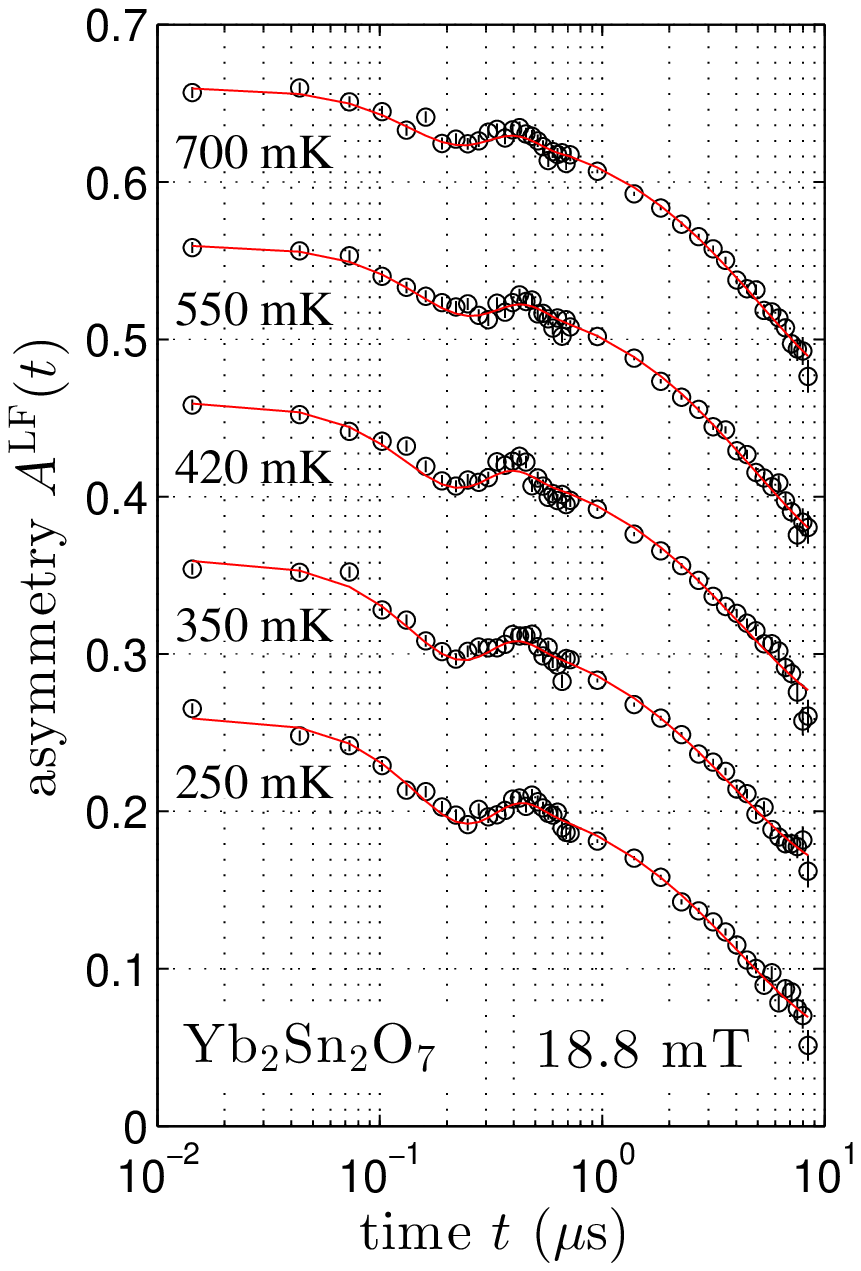}
\includegraphics[width=0.5\linewidth ]{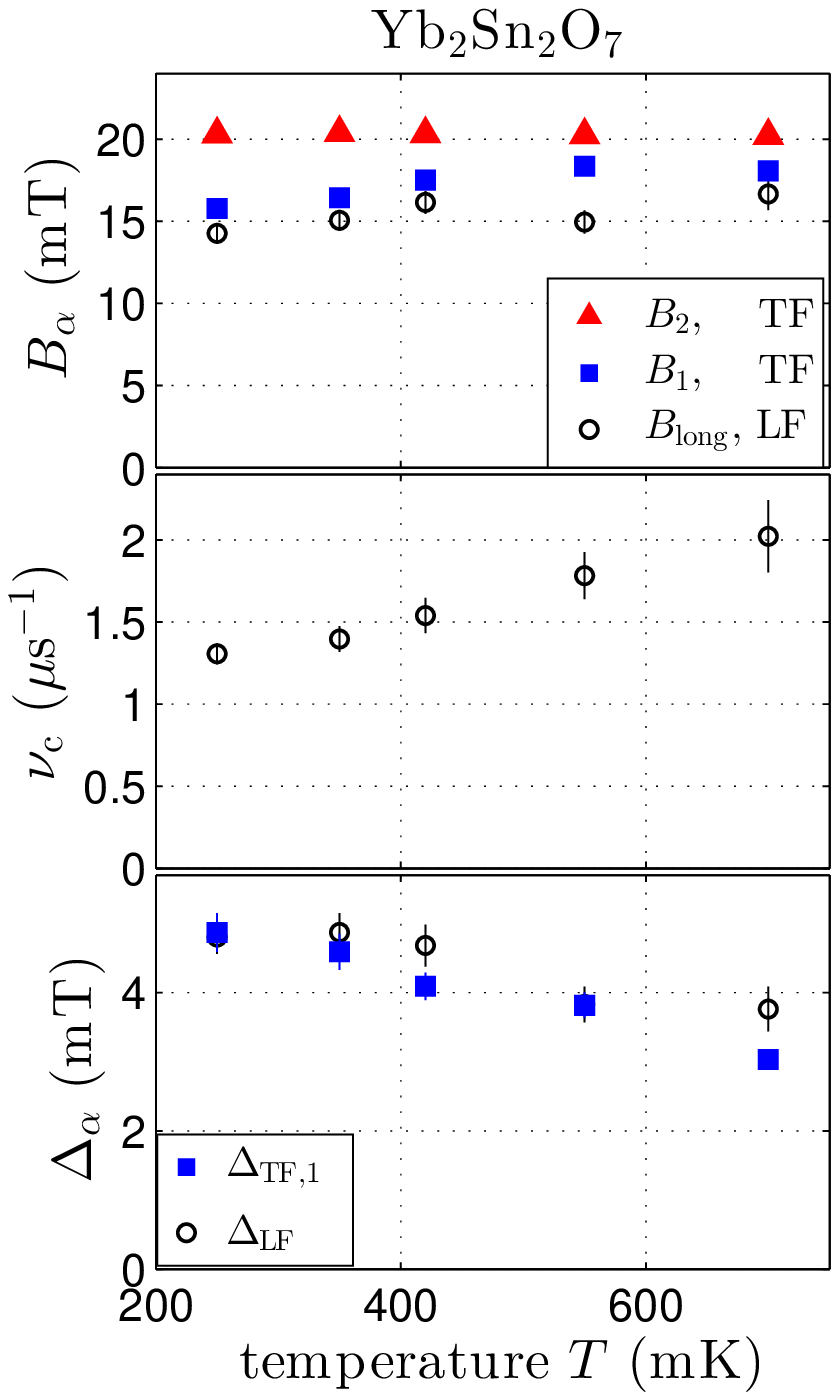}  
\hspace{-0.8cm}
\caption{(color online). 
Results obtained in the paramagnetic phase of a powder sample of Yb$_2$Sn$_2$O$_7$ under $B_{\rm ext} = 18.8$~mT.
(left) Examples of asymmetry spectra obtained at LTF in the longitudinal geometry. Circles are experimental 
data while solid lines represent fits as explained in the main text. The temperature is given 
nearby each of the spectra which are vertically shifted by 0.1 units for better visualization.
(right) Parameters extracted from a fit to Eqs.~\ref{eq:asyStatDef}--\ref{eq:AsyModel:LF} for LF data and Eq.~\ref{eq:TFmodel} for TF data.
}
\label{fig:YbSn2Ti2:Tdep}
\end{figure}

In Fig.~\ref{fig:YbSn2Ti2:Tdep} we display results 
obtained in a LF of $B_{\rm ext}$ = 18.8~mT above $T_{\rm c}=0.15$ K. Also in this compound one can 
clearly see oscillations in the asymmetry spectra associated with $B_{\rm ext}$, which directly point to a slow spin dynamics. The dynamical Gaussian Kubo-Toyabe model, 
i.e.\ Eqs.~\ref{eq:asyStatDef}--\ref{eq:AsyModel:LF} with $\eta_3=\eta_4=0$ provides a good account of the spectra. 
The local mean field, i.e.\ $B_{\rm long}$, in this compound is again reduced. The strength of the reduction 
is weaker than in Yb$_2$Ti$_2$O$_7$ and it also decreases with increasing temperature. 
The difference between $B_1$ and $B_{\rm long}$
is small and may not be really significant. The spin fluctuation rate $\nu_{\rm c}$ gradually 
increases with temperature and reaches a maximal value of $2.0 \, (2)~\mu{\rm s}^{-1}$ at 700~mK. 
As expected, the magnitudes of $\Delta_{{\rm TF},1}$ and $\Delta_{\rm LF}$ are almost equal. This is in contrast to 
Yb$_2$Ti$_2$O$_7$. 

\begin{figure}[!t]
\hspace{-0.8cm}
\includegraphics[width=0.5\linewidth]{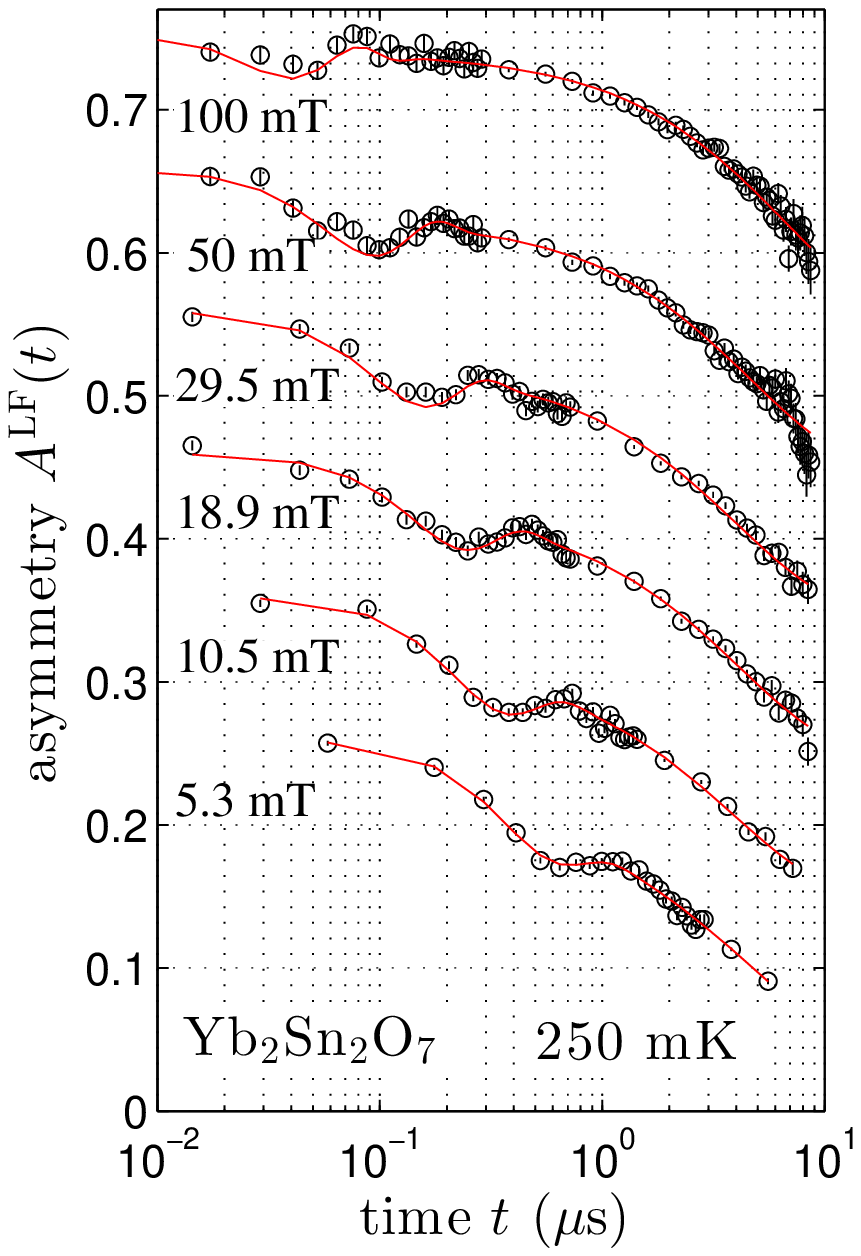}
\includegraphics[width=0.5\linewidth]{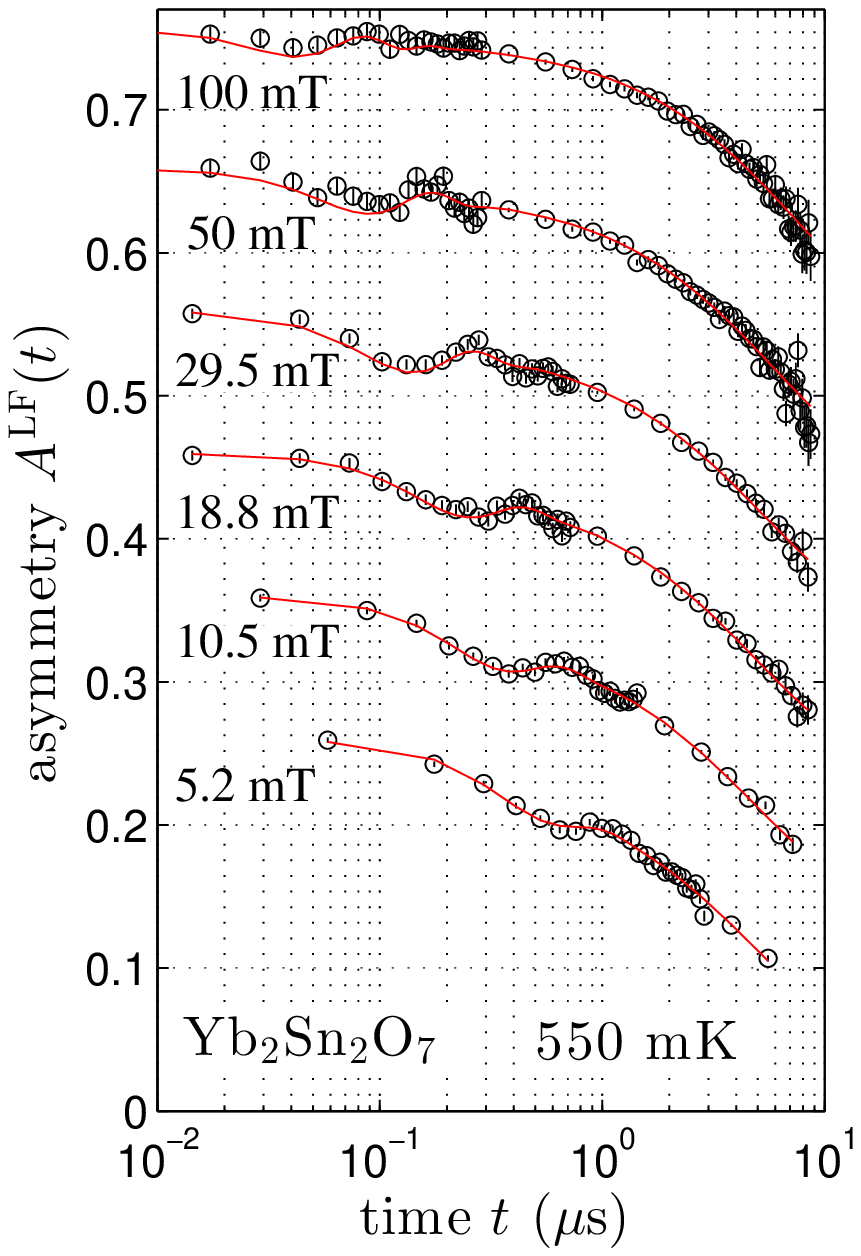}
\hspace{-0.8cm}
\caption{(color online). Examples of field scan measurements of asymmetry spectra performed in 
the paramagnetic phase of a powder sample of Yb$_2$Sn$_2$O$_7$ (black circles) and corresponding 
fitting curves obtained with Eqs.~\ref{eq:asyStatDef}--\ref{eq:AsyModel:LF}. 
The spectra were recorded in the longitudinal geometry at the LTF spectrometer for $T = 250$~mK  
(left) and  $T= 550$~mK  (right). The $B_{\rm ext}$ value is 
given nearby the spectra which are shifted by 0.1 unit relative to each other for better visualization. 
It is remarkable that oscillations due to $B_{\rm ext}$ are clearly observed  
pointing to a low frequency spin dynamics above $T_{\rm c}$ in Yb$_2$Sn$_2$O$_7$. 
}
\label{fig:Yb2Ti2O7:Bdep:Spectra}
\end{figure}

\begin{figure}[t]
\centering
\includegraphics[width=0.5\linewidth]{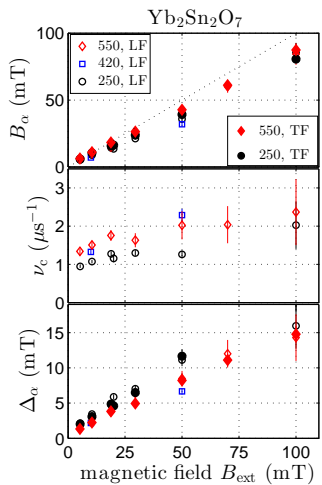}
\caption{(color online). 
Same caption as for Fig.~\ref{Yb2Ti2O7_para_parameters}. Here the physical parameters concern the paramagnetic phase of Yb$_2$Sn$_2$O$_7$. 
}
\label{fig:Yb2Ti2O7:Bdep:Params}
\end{figure}

In order to clarify the origin of the slow spin dynamics in the paramagnetic state additional measurements at 250, 420, and 550~mK were performed for different fields. In 
Fig.~\ref{fig:Yb2Ti2O7:Bdep:Spectra} 
we show examples at 250 and 550~mK, together with the fitting curves depicted as solid 
lines. The oscillations due to $B_{\rm ext}$ are present in the whole field range of the measurements. 
Again, a Gaussian field component distribution is sufficient for the description of the spectra. In Fig.~\ref{fig:Yb2Ti2O7:Bdep:Params} 
the corresponding fitting parameters are presented as a function of field. The field dependence of the three 
parameters measured for the stannate and the titanate in their paramagnetic states is rather similar. 

\subsection{Ordered magnetic state} 
\label{Yb2Sn2O7_data_order}

\begin{figure}[!t] 
\begin{picture}(255,300)
\put(0,110){\includegraphics[width=0.49\linewidth]{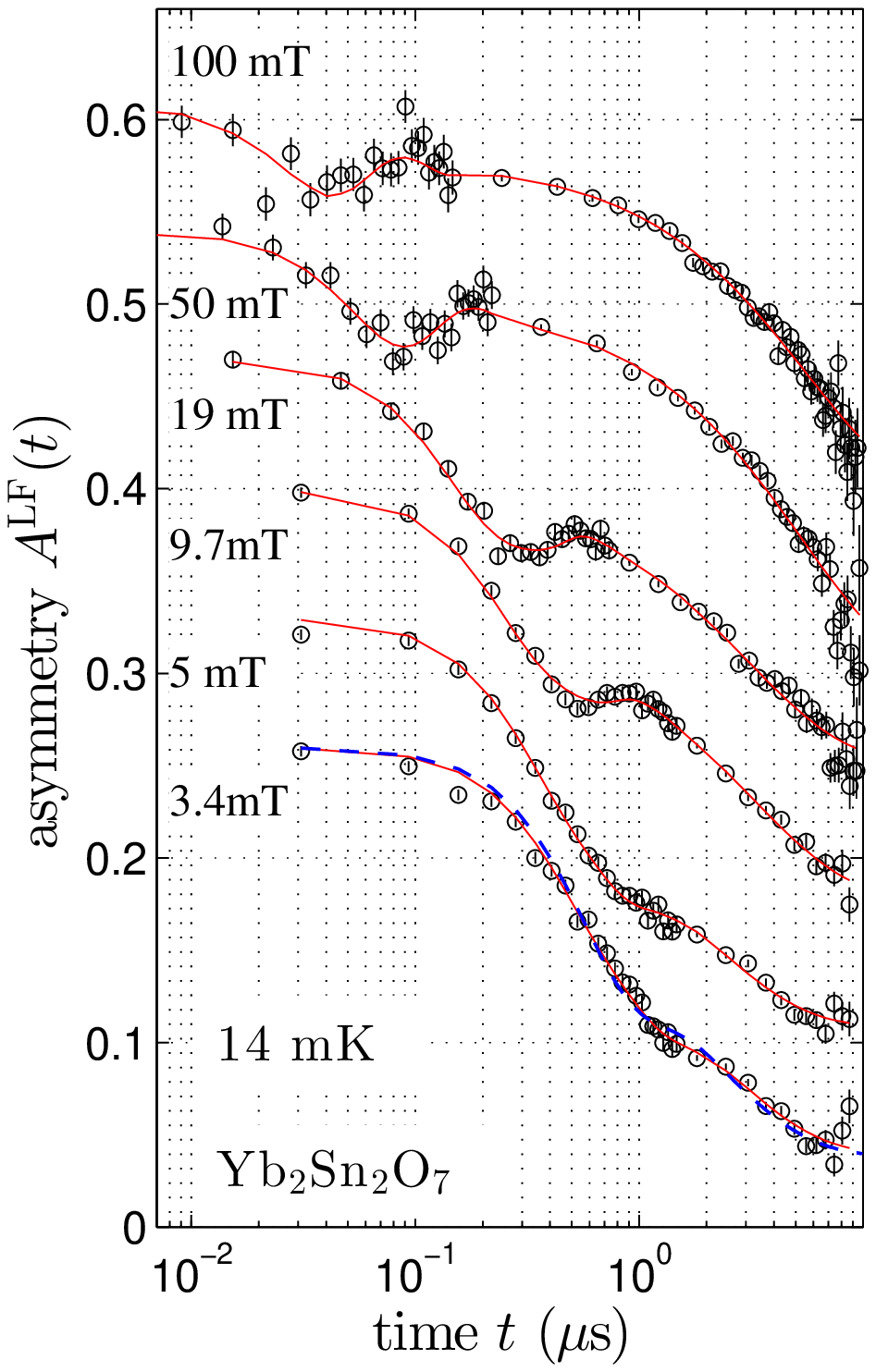}}
\put(88,138){\small(a)}
\vspace{3.58cm}
\put(-10,-10){\includegraphics[width=0.59\linewidth]{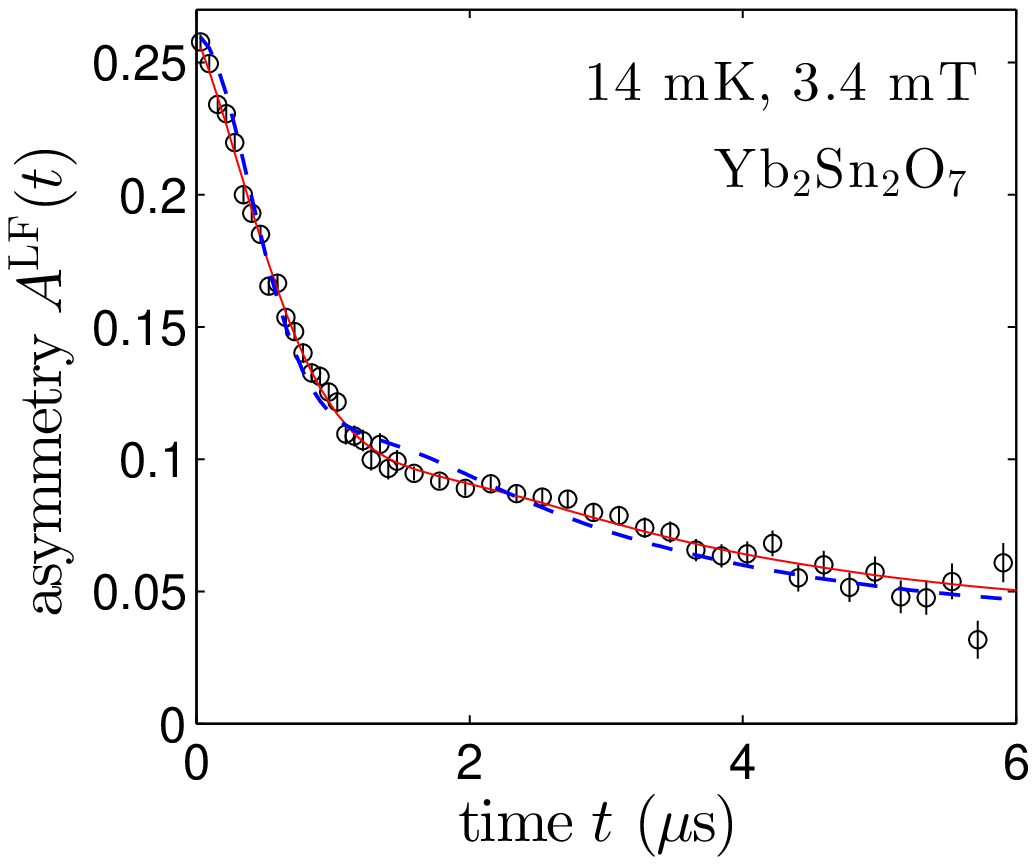}}
\put(98,70){\small(e)}
\put(127,110){\includegraphics[width=0.50\linewidth]{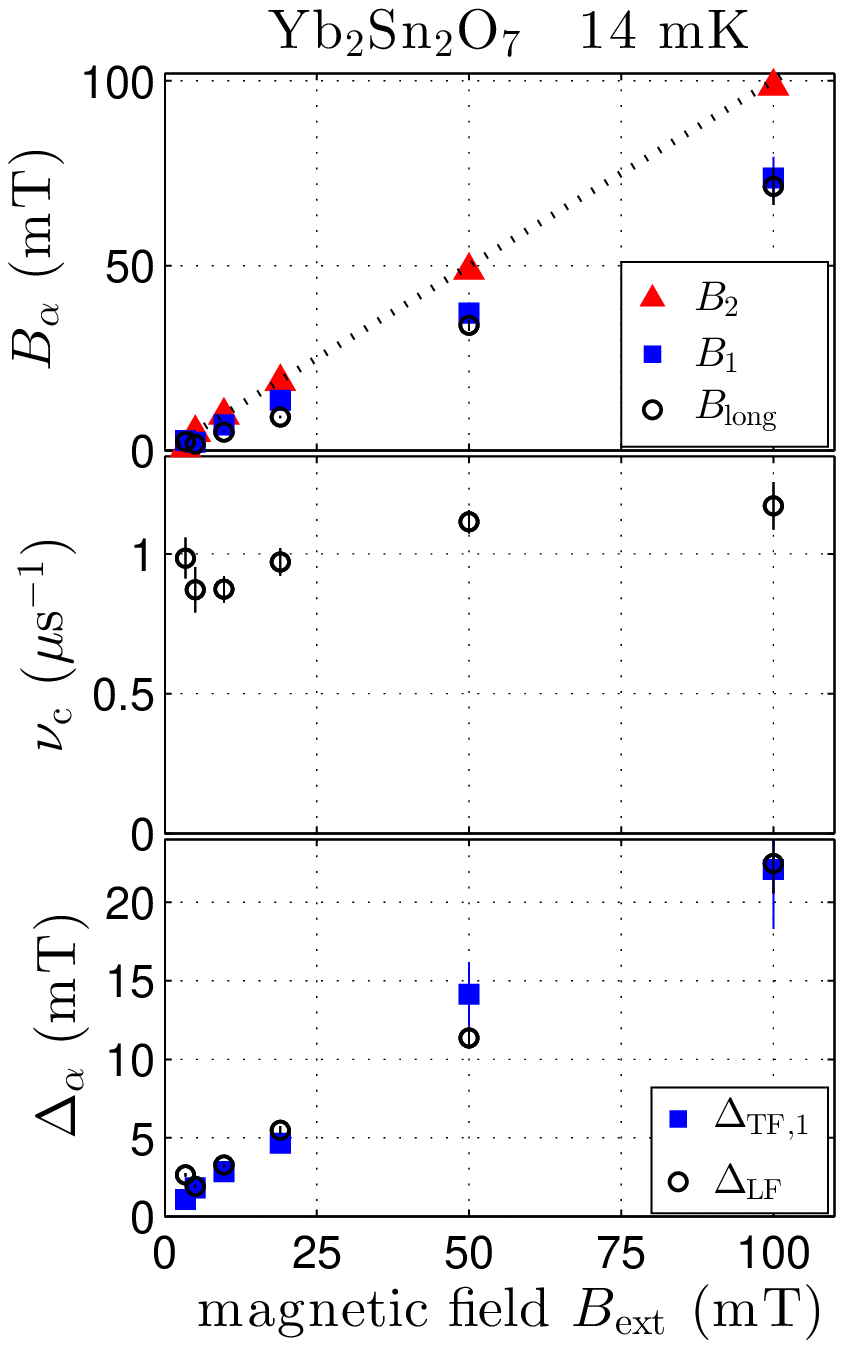}}
\put(132,0){\includegraphics[width=0.48\linewidth]{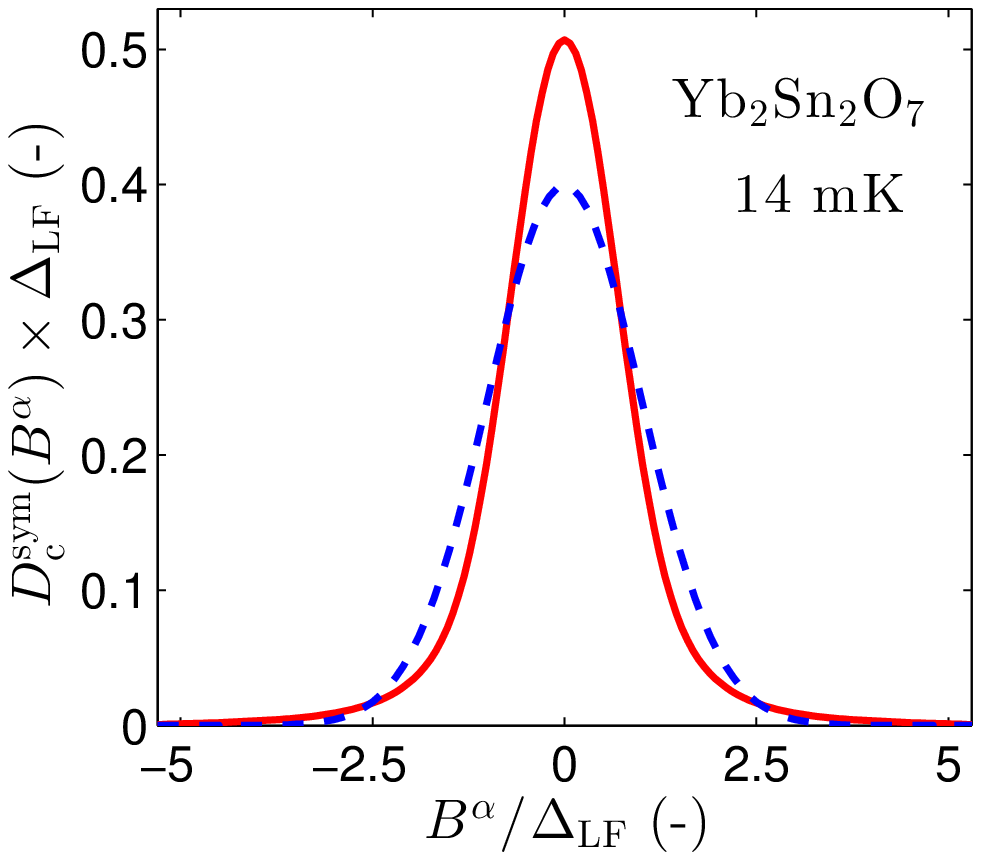}}
\put(225,65){\small(f)}
\put(155,280){\small(b)}
\put(155,192){\small(c)}
\put(155,170){\small(d)}
\end{picture}
\caption{(color online).
Data recorded at the LTF spectrometer for a powder sample of Yb$_2$Sn$_2$O$_7$ at $T = 14$~mK with the longitudinal geometry. (a) 
Field dependence of LF time spectra (circles) and fits (solid lines) as described in the main text. The spectra are vertically shifted by 0.07 for better visualization (b, c, d) Corresponding fitting parameters as a function of $B_{\rm ext}$. The field component standard deviation is $\Delta_{\rm LF}=1.38 \, \delta$ for $\eta_3=0.670$ and $\eta_4=0.400$. The $B_{\rm ext}$ dependence of the parameters deduced from the TF spectra is also displayed.  
(e) LF time spectrum (circles) at $3.4$~mT plotted with a linear time scale and fits with the 
extension of the Gaussian model in red solid line and the conventional Gaussian Kubo-Toyabe model in blue 
dashed line. (f) Field component distributions in normalized field units. The solid line shows the 
distribution obtained from a global fit using the extension of the Gaussian model, and the dashed line 
the Gaussian distribution corresponding to the Kubo-Toyabe fit in panel (e), also in blue 
dashed line.
}
\label{fig:Yb2Sn2O7:Bdep:Magn:Spectra}
\end{figure}

Spectra were recorded in the magnetically ordered state at $14$~mK. They were presented in
Ref.~\onlinecite{Yaouanc13}. The sample was first zero-field cooled and then ${\bf B}_{\rm ext}$ was applied 
with the following sequence: $5$, $3.4$, $9.7$, $19$, $50$, and $100$~mT. The LF and TF asymmetry spectra 
are well described with Eqs.~\ref{eq:asyStatDef}--\ref{eq:AsyModel:LF} and Eq.~\ref{eq:TFmodel}, respectively. 
As for  Yb$_2$Ti$_2$O$_7$ the parameters 
$\eta_3$ and $\eta_4$ are appreciable and field independent within the statistical uncertainties. Therefore, 
we performed a global fit of the LF spectra with common $\eta_3$ and $\eta_4$ in order to determine the field distribution. The experimental LF time asymmetries together with the fitting curves are presented in Fig.~\ref{fig:Yb2Sn2O7:Bdep:Magn:Spectra}(a). 
The analysis results in $\eta_3 = 0.67\, (4)$ and $\eta_4=0.40 \, (3)$, i.e.\ only slightly
different from the values measured for Yb$_2$Ti$_2$O$_7$. From these values we get $\Delta_{\rm LF}$ = 1.38\,(9)\,$\delta$. In 
Figs.~\ref{fig:Yb2Sn2O7:Bdep:Magn:Spectra}(b)--\ref{fig:Yb2Sn2O7:Bdep:Magn:Spectra}(d)
we show the field dependence of the other LF parameters, together with the TF parameters. 
A linear dependence of the field standard deviations is also present 
in the ordered state. The fluctuation rate $\nu_{\rm c}$ is slightly reduced and exhibits a weaker 
field dependence than in the paramagnetic state. The fields $B_{\rm long}$ and $B_1$ are equal within 
experimental uncertainties. They are reduced relative to $B_{\rm ext}$, but not as strongly as in the case
of Yb$_2$Ti$_2$O$_7$; see Fig.~\ref{Yb2Ti2O7_ordered_Fscan_spectra}.
In Fig.~\ref{fig:Yb2Sn2O7:Bdep:Magn:Spectra}(e) we compare the results of the fits with the Gaussian model and with its extension. The latter solution provides a better description. The field component distribution 
obtained from this analysis is shown in  Fig.~\ref{fig:Yb2Sn2O7:Bdep:Magn:Spectra}(f) as a solid line.  As for 
Yb$_2$Ti$_2$O$_7$, a significant deviation from the Gaussian shape displayed as a dashed line is evident: pronounced
tails are present. 

\section{Discussion} 
\label{Discussion}

We shall first consider the conclusions reached here from the analysis of spectra recorded under longitudinal
fields. Then we shall discuss their compatibility with those previously derived from extremely low field measurements. 

\subsection{Measurements under relatively large fields} 
\label{Discussion_LF}

Let us first summarize the results of the analysis of the $\mu$SR spectra for the two compounds discussed in 
this report, i.e.\ Yb$_2$Ti$_2$O$_7$ and Yb$_2$Sn$_2$O$_7$. 

The following features emerge. From the highest 
temperature investigated, i.e.\ $700$~mK, down to far below the temperature $T_{\rm c}$ at which a sharp peak in 
the specific heat is detected, the fluctuation rate of the Yb$^{3+}$ magnetic moments  
is equal to $\nu_{\rm c} \approx 1 \, \mu {\rm s}^{-1}$, independent of the field and temperature within a factor  
$2$. Hence, the spin dynamics is anomalously slow. The component field distribution measured under longitudinal 
field is Gaussian above $T_{\rm c}$ and somewhat deviates from it below 
$T_{\rm c}$. Its standard deviation increases linearly with the field for the two compounds, both below and above $T_{\rm c}$. When normalized to the width, the distribution is field independent in the probed field range. Finally, 
a reduction of the mean field at the muon site is measured for the two compounds. Within experimental 
uncertainties, it is similar in longitudinal and transverse field measurements. It is particularly 
strong for Yb$_2$Ti$_2$O$_7$ at $T< T_{\rm c }$. Remarkably, it is approximately linear in field, both below and 
above $T_{\rm c}$. The ratio $(B_\alpha-B_{\rm ext})/B_{\rm ext}$ can therefore be considered as a local magnetic susceptibility.

As the magnetic phase transitions are first order, the procedure followed to record spectra 
below $T_{\rm c}$ has an influence on the results of the measurements. We recall that the asymmetry spectra 
analyzed  in this report were taken after cooling the samples in zero field, except for the data of 
Fig.~\ref{Yb2Ti2O7_FTexample:ordered}. 

We now discuss the meaning of our results. 
We first note that the existence of slow spin dynamics in magnetic pyrochlore compounds is well documented. It was 
reported in 2002 
for Yb$_2$Ti$_2$O$_7$ from the analysis of $\mu$SR and M\"ossbauer measurements.\cite{Hodges02} Soon after,
it was realized that a wide spectrum of fluctuation rates exists in this compound.\cite{Gardner04} Later on,
slow spin dynamics was reported for the pyrochlore Tb$_2$Ti$_2$O$_7$ using ac susceptibility 
measurements.\cite{Ueland06} Slow spin dynamics with a wide range of fluctuation rates of the rare-earth magnetic 
moments in geometrically frustrated magnetic compounds is in fact one of their characteristics. As an additional 
example, we mention the garnet compound Gd$_3$Ga$_5$O$_{12}$ with three-dimensional hyperkagom\'e magnetic lattice for which neutron-scattering 
studies reveal a remarkable wide range of energies from milli- to picoelectronvolt.\cite{Deen10} 
The present study puts on firm ground the existence of slow dynamics for Yb$_2$Ti$_2$O$_7$ and Yb$_2$Sn$_2$O$_7$
as seen by $\mu$SR.

As to the physical mechanism at the origin of the slow dynamics, we note that 
persistent spin dynamics at low temperature, i.e.\ below $\simeq 1$~K, as observed by $\mu$SR spin-lattice relaxation
in pyrochlore compounds has recently been proposed to arise from unidimensional excitations that would be supported by 
spin loops.\cite{Yaouanc15} These one-dimensional objects in three-dimensional crystal structures such as the pyrochlore 
structure have been studied early on.\cite{Villain79,Hermele04} Their existence provides a natural explanation for 
slow dynamics since it means that spin loops, i.e.\ a large number of spins in contrast to a conventional 
single spin, are involved in 
the dynamics. This picture may extend above $\simeq 1$~K, the interaction between spin loops being recognized.
This is in qualitative agreement with the observation of slow spin dynamics deep into the paramagnetic state of 
Nd$_2$Sn$_2$O$_7$.\cite{Bertin15} Interestingly, classical simulations of spin dynamics for the antiferromagnetic 
Heisenberg model on the kagom\'e lattice is consistent with the loop picture; see Ref.~\onlinecite{Taillefumier14} 
and references therein.

The field component distribution extracted from the measurements below $T_{\rm c}$ deviates from the Gaussian
shape for both compounds, pointing out the existence of short-range correlations. These correlations are
absent in the ordered magnetic state of a conventional compound. They do exist for Er$_2$Ti$_2$O$_7$.\cite{Ruff08}
The increase of the field standard deviation with $B_{\rm ext}$ 
suggests that the field induces some disorder. The mechanism at play is 
not clear. It could be that the field induces a canting of the magnetic moments. Surprisingly, we have not found any effect of magnetic short-range correlations on the field component distribution in the paramagnetic state. Yet such correlations, signalled by a broad hump in the specific heat centered at about $3$ and $2$~K for Yb$_2$Ti$_2$O$_7$ and Yb$_2$Sn$_2$O$_7$, respectively, do exist\cite{Blote69,Yaouanc11c,Yaouanc13,Dalmas03} and have even been observed by neutron scattering\cite{Bonville03a,Bonville04a} in a crystal of Yb$_2$Ti$_2$O$_7$ whose specific heat does nonetheless display no sharp peak at $T_{\rm c}$.\cite{Yaouanc11c} The detection of these correlations by $\mu$SR needs to be further investigated. 

Let us now discuss the origin for the reduction of the mean field at the muon site. We have the key result 
that the same reduction is measured for LF and TF asymmetries. This suggests that in fact we are dealing with 
a frequency shift effect and not an effect of intermittency of the field at the muon site.\cite{Uemura94} This interpretation is supported by the results published for 
Tb$_2$Ti$_2$O$_7$.\cite{Yaouanc11a} In this system, below 10~K, the relative frequency shift varied between $-0.2$ and $-0.7$. In Yb$_2$Ti$_2$O$_7$ [Fig.~\ref{Yb2Ti2O7_ordered_Fscan_spectra}(b)] it is approximately $-0.6$ and for Yb$_2$Sn$_2$O$_7$ [Fig.~\ref{fig:Yb2Sn2O7:Bdep:Magn:Spectra}(b)] it is near $-0.2$. In conclusion, there is no exotic physics in the measured field 
reduction. It just occurs that the frequency shift is negative. If the muon site 
localization were known, it would probably be possible to understand this feature. This has recently been done for
MnSi.\cite{Amato14} The much larger reduction below $T_{\rm c}$ for Yb$_2$Ti$_2$O$_7$ relative to Yb$_2$Sn$_2$O$_7$
is just the signature of a much larger local magnetic susceptibility for the former compound.

\subsection{Compatibility with previously published extremely low field measurements} 
\label{Discussion_ZF}

The asymmetry $A^{\rm LF}(t)$ and therefore the polarization function $P_Z(t)$ in the paramagnetic state does not decay 
monotonically at intermediate $B_{\rm ext}$ values for Yb$_2$Ti$_2$O$_7$ and Yb$_2$Sn$_2$O$_7$. On the other hand, at extremely 
low fields it does.\cite{Hodges02,Yaouanc13} Quantum perturbation theory describes $P_Z(t)$ in terms of 
correlation functions when the spin dynamics is sufficiently fast,\cite{Baryshevskii76,McMullen78,Hayano79,Dalmas92}
namely $\nu_{\rm c} /(\gamma_\mu \Delta_{\rm LF}) \gtrsim 2$.\cite{Yaouanc11} It has been shown that this quantum description 
and the one exposed in Sec.~\ref{Experimental_data_analysis} give consistent results in this limit when the field component
distribution is Gaussian,\cite{Yaouanc11}  as in the paramagnetic state of the two compounds of interest. Remarkably, at 
extremely low $B_{\rm ext}$ the quantum description leads to a monotonic decay of $P_Z(t)$ which is exponential-like
for $\nu_{\rm c}t  \gtrsim 1$.\cite{Dalmas92,Keren94} In our case, the shortest time we probed is $t \simeq$~0.1~$\mu$s.
This means that $\nu_{\rm c} \gtrsim 10$~$\mu$s$^{-1}$, a lower bound larger than the value extracted from the analysis 
of $A^{\rm LF}(t)$ measured under field which is $\nu_{\rm c} \approx$~1~$\mu$s$^{-1}$. Hence, the fluctuation modes probed 
at extremely low field are fluctuating much faster than at intermediate field. 

This discussion gives results entirely consistent with previous inferences from extremely low $B_{\rm ext}$ measurements near $T_{\rm c}$ in the paramagnetic state.\cite{Hodges02,Yaouanc13} From $^{170}$Yb M\"ossbauer measurements the fluctuation rate for Yb$_2$Ti$_2$O$_7$ and Yb$_2$Sn$_2$O$_7$ was found to be in the $10^{10}$~s$^{-1}$ range. Identifying $\mu$SR and M\"ossbauer fluctuation rates, taking into account the measured spin-lattice relaxation rate $\lambda_Z$ characterizing the decay of $P_Z(t)$, and the conventional motional narrowing formula $\lambda_Z = 2 \gamma^2_\mu \Delta^2_{\rm LF}/\nu_{\rm c}$, $\Delta_{\rm LF}$ was computed
to be consistent with expectation for the known Yb$^{3+}$ magnetic moments. 

At this juncture we note that the field standard deviation
related to the slow dynamics is more than one order of magnitude smaller than the one characterizing the faster dynamics. This means 
that only a small fraction of the Yb$^{3+}$ magnetic modes is involved in the slow dynamics.     

Surprisingly, no spontaneous $\mu$SR oscillation is observed below $T_{\rm c}$ for the stannate. Such a lack of oscillation was also observed for Tb$_2$Sn$_2$O$_7$ and interpreted as a signature of 
the dynamical nature of the ground state of the compound,\cite{Dalmas06} later on confirmed by neutron-spin echo 
measurements.\cite{Rule09b} This means that 
${\tilde \nu_{\rm c}} > 2\gamma_\mu B_{\rm fluc}$, where ${\tilde \nu_{\rm c}}$ is the fluctuation rate of the fluctuating spontaneous
field $B_{\rm fluc}$ at the muon site.\cite{Yaouanc11} From the relaxation rate $\lambda_Z\simeq$~2.3~$\mu$s$^{-1}$ measured at low temperature in Tb$_2$Sn$_2$O$_7$ (Ref.~\onlinecite{Dalmas06}) and from the fluctuation rate $\tilde \nu_{\rm c}$ = $5\times 10^{10}$~s$^{-1}$ measured with the neutron spin echo technique, we deduce $B_{\rm fluc}$ = 0.4~T using the formula $\lambda_Z = \gamma_\mu^2 B_{\rm fluc}^2/\nu_{\rm c}$. Scaling $B_{\rm fluc}$ to the ordered magnetic moments --- 5.4\,$\mu_{\rm B}$ in the Tb$_2$Sn$_2$O$_7$ and approximately 1.1\,$\mu_{\rm B}$ in the Yb pyrochlores\cite{Hodges02,Yaouanc13} --- we compute 
$B_{\rm fluc} \simeq (1.1/5.4) \times 0.4 = 0.08$~T for the two ytterbium compounds. Therefore ${\tilde \nu_{\rm c}} > 7 \times 10^7$~s$^{-1}$. 

Hence, the analysis of the $\mu$SR data leads to two magnetic fluctuation rates in the paramagnetic state: 
$\approx 10^6$~s$^{-1}$  and $\approx 10^{10}$~s$^{-1}$ close to $T_{\rm c}$. In the ordered state two rates are also detected: again $\approx 10^6$~s$^{-1}$ 
and a rate larger than $7\times 10^7$~s$^{-1}$. Finding more than one fluctuation mode is not surprising: we refer to the ordered 
state of Tb$_2$Sn$_2$O$_7$.\cite{Dalmas06,Chapuis07,Mirebeau08,Rule09b}

\section{Conclusions} 
\label{Conclusions}

In this paper we have presented a complete $\mu$SR study of the spin dynamics of Yb$_2$Ti$_2$O$_7$ and 
Yb$_2$Ti$_2$O$_7$ as seen in a modest applied field. Extremely slow fluctuation modes with a scale of approximately $10^6$~s$^{-1}$ were found both below and above the transition temperature. On the other hand, a much faster fluctuation rate, namely in the 
$10^{10}$~s$^{-1}$ range near $T_{\rm c}$, was recognized from extremely low field paramagnetic measurements. In addition, a scale larger than $7\times 10^7$~s$^{-1}$ does exist in the ordered state. A similar spectrum of fluctuation rates is present in the two compounds. Such a broad and complex spectrum of fluctuations is unusual in a magnetic system. To our knowledge no theory is available for comparison.

The field component distribution deviates from the Gaussian shape for the two compounds 
below $T_{\rm c}$, pointing out the existence of short-range magnetic correlations. This is an unconventional
feature which suggests the formation of molecular spin substructures, such as spin loops. They probably also exist
in the paramagnetic state, although a signature of them has only been found from specific heat. Finally our study reveals that the magnetic susceptibility of Yb$_2$Ti$_2$O$_7$ is much stronger than that of Yb$_2$Sn$_2$O$_7$ below $T_{\rm c}$.

Still considering the ytterbium pyrochlore family, we note that ferromagnetic superexchange interactions drive the 
magnetic transition observed at $T_{\rm c}$ for Yb$_2$Ti$_2$O$_7$ and Yb$_2$Sn$_2$O$_7$. These compounds occur to
be spin-ice like. It would be of interest to investigate with $\mu$SR techniques Yb$_2$Ge$_2$O$_7$ for which 
antiferromagnetic superexchange interactions are responsible for the magnetic transition.\cite{Dun14} In 
particular, possible differences in the spin dynamics would be of interest. In this respect, we note that the
relaxation for the antiferromagnet Er$_2$Ti$_2$O$_7$ is strikingly different from observed in this 
report.\cite{Lago05,Dalmas12a} In particular, the field responses below $T_{\rm c}$ are quite 
different.

\section*{Acknowledgments}

This research project has been partially supported by the European Commission under the 
6th Framework Programme through the Key Action: Strengthening the European Research 
Area, Research Infrastructures (Contract number: RII3-CT-2003-505925) and 
under the 7th Framework Programme through the `Research Infrastructures' action of the
`Capacities' Programme, Contract No: CP-CSA\_INFRA-2008-1.1.1 Number
226507-NMI3. The $\mu$SR measurements were performed at S$\mu$S, Paul Scherrer Institute, 
Villigen, Switzerland, and at the ISIS facility, Rutherford Appleton Laboratory, 
Chilton, UK. AY gratefully acknowledges partial support of Prof.\ H.\ Keller from the 
University of Zurich for the $\mu$SR measurements.

\bibliography{reference.bib}

\end{document}